\documentclass[journal]{IEEEtran}
\IEEEoverridecommandlockouts
\usepackage{cite}
\usepackage{amsmath,amssymb,amsfonts}
\usepackage{algorithmic}
\usepackage{graphicx}
\usepackage{textcomp}
\usepackage{xcolor}
\usepackage{amsfonts}
\usepackage{bm}
\usepackage{booktabs}
\usepackage{algorithm}
\usepackage{multirow,bm,bbm,array,setspace}
\usepackage{textcomp}
\usepackage{psfrag}
\usepackage{pstricks,enumerate}
\usepackage{bbm,subfigure}
\usepackage{dblfloatfix}
\usepackage{xpatch}

\makeatletter
\newcommand{\distas}[1]{\mathbin{\overset{#1}{\kern\z@\sim}}}%
\newsavebox{\mybox}\newsavebox{\mysim}
\newcommand{\distras}[1]{%
  \savebox{\mybox}{\hbox{\kern1pt$\scriptstyle#1$\kern1pt}}%
  \savebox{\mysim}{\hbox{$\sim$}}%
  \mathbin{\overset{#1}{\kern\z@\resizebox{\wd\mybox}{\ht\mysim}{$\sim$}}}%
}

\ExplSyntaxOn
\cs_new:Npn \bibColoredItems #1#2
  {
    \clist_map_inline:nn {#2} { \cs_new:cpn {bib@colored@##1} {#1} } 
  }
\ExplSyntaxOff

\newcommand\bib@setcolor[1]{%
  \ifcsname bib@colored@#1\endcsname
    \expandafter\color\expandafter{\csname bib@colored@#1\endcsname}
  \else
    \normalcolor
  \fi
}

\xpatchcmd\@bibitem
  {\item}
  {\bib@setcolor{#1}\item}
  {}{\PatchFailed}

\xpatchcmd\@lbibitem
  {\item}
  {\bib@setcolor{#2}\item}
 {}{\PatchFailed}
\makeatother
\ifCLASSINFOpdf
\else
\fi

\newcommand{\bA}{\bm A}
\newcommand{\bZ}{\bm Z}
\newcommand{\bC}{\bm C}

\newcommand{\bB}{\bm B}
\newcommand{\bM}{\bm M}
\newcommand{\bX}{\bm X}

\newcommand{\bY}{\bm Y}

\newcommand{\bfA}{\bm A}
\newcommand{\bfB}{\bm B}
\newcommand{\bfC}{\bm C}
\newcommand{\bfM}{\bm M}
\newcommand{\bW}{\bm W}

\newcommand{\bL}{\bm L}
\newcommand{\bfL}{\bm L}
\newcommand{\bfK}{\bm K}

\newcommand{\bfD}{\bm D}

\newcommand{\bG}{\bm G}

\newcommand{\bI}{\bm{I} }
\newcommand{\bfa}{\bm a}
\newcommand{\bfY}{\bm Y}

\newcommand{\bfV}{\bm V}
\newcommand{\bhh}{\text {H}}
\newcommand{\ttr}{\mathrm {tr}}
\newcommand{\tr}{\mathrm {tr}}

\newcommand{\bfF}{\bm F}

\newcommand{\bfs}{\bm s}
\newcommand{\bfX}{\bm X}
\newcommand{\bfZ}{\bm Z}

\newcommand{\bfS}{\bm S}
\newcommand{\bfW}{\bm W}
\newcommand{\bfH}{\bm H}
\newcommand{\bfG}{\bm G}
\newcommand{\bfQ}{\bm Q}
\newcommand{\bfU}{\bm U}

\newcommand{\bGa}{\bm\Gamma}

\newtheorem{proposition}{Proposition}
\newtheorem{lemma}{Lemma}

\newtheorem{remark}{Remark}
\newtheorem{example}{Example}

\begin{document}

\title{Fast Fractional Programming for Multi-Cell Integrated Sensing and Communications}
\author{
	\IEEEauthorblockN{
	Yannan Chen, \IEEEmembership{Graduate Student Member,~IEEE}, Yi Feng, \IEEEmembership{Graduate Student Member,~IEEE},\\Xiaoyang Li, \IEEEmembership{Member,~IEEE},
 Licheng Zhao, Kaiming Shen, \IEEEmembership{Senior Member,~IEEE}
} 
\thanks{Manuscript accepted to IEEE Transactions on Wireless Communications on March 19, 2025. This work was supported in part by the NSFC under Grant 12426306 and Grant 62206182, in part by Guangdong Basic and Applied Basic Research under Grant 2023B0303000001 and Grant 2024A1515010154, in part by Shenzhen Steady Funding Program, and in part by Shenzhen Science and Technology Program under Grant JCYJ20220530113017039 and Grant JCYJ20241202124934046. An earlier version of this paper was presented in part
at the International Conference on Wireless Communications and Signal Processing (WCSP) [10.1109/WCSP62071.2024.10827681], Hefei, China, 2024. \emph{(Corresponding author: Kaiming Shen.)}

Yannan Chen, Yi Feng, and Kaiming Shen are with School of Science and Engineering, The Chinese University of Hong Kong (Shenzhen), Shenzhen 518172, China (e-mail: yannanchen@link.cuhk.edu.cn; yifeng@link.cuhk.edu.cn; shenkaiming@cuhk.edu.cn).

Xiaoyang Li and Licheng Zhao are with Shenzhen Research Institute of Big Data, The Chinese University of Hong Kong (Shenzhen), Shenzhen 518172, China (e-mail: lixiaoyang@sribd.cn; zhaolicheng@sribd.cn).
}
}
\maketitle

\begin{abstract}
This paper concerns the coordinate multi-cell beamforming design for integrated sensing and communications (ISAC). In particular, we assume that each base station (BS) has massive antennas. The optimization objective is to maximize a weighted sum of the data rates (for communications) and the Fisher information (for sensing). We first show that the conventional beamforming method for the multiple-input multiple-output (MIMO) transmission, i.e., the weighted minimum mean square error (WMMSE) algorithm, works for the ISAC problem case from a fractional programming (FP) perspective. However, the WMMSE algorithm frequently requires computing the $N\times N$ matrix inverse, where $N$ is the number of transmit or receive antennas, so the algorithm becomes quite costly when antennas are massively deployed. To address this issue, we develop a nonhomogeneous bound and use it in conjunction with the FP technique to solve the ISAC beamforming problem without the need to invert any large matrices. It is further shown that the resulting new FP algorithm has an intimate connection with gradient projection, based on which we can accelerate the convergence via Nesterov's gradient extrapolation.

\end{abstract}

\begin{IEEEkeywords}
Integrated sensing and communication (ISAC), multi-cell beamforming, massive antennas, large matrix inverse, convergence acceleration.
\end{IEEEkeywords}

\section{Introduction}
\IEEEPARstart{I}{NTEGRATED} sensing and communications (ISAC) is an emerging wireless technique that reuses the network infrastructure and radio signals for both communications and sensing---which used to be dealt with separately in the conventional networks, in order to reduce the infrastructure cost and boost the spectral efficiency. This work focuses on the large antenna array case of ISAC, aiming at a system-level optimization by coordinating the antenna beamformers across multiple cells. There are three main results worth noticing about this work. First, we show that the well-known weighted minimum mean square error (WMMSE) algorithm \cite{cioffi_WMMSE,shi2011iteratively}, which was initially devised for the multiple-input multiple-output (MIMO) transmission, can be applied for ISAC from a fractional programming (FP) perspective, but at a high computation cost due to the large matrix inverse operation. Second, we manage to eliminate all the large matrix inversion from the conventional FP by incorporating into FP a generalized nonhomogeneous bound \cite{sun2016majorization}. Third, the above method turns out to have an intimate connection with gradient projection, so Nesterov's extrapolation scheme \cite{Nesterov_book} can be used to accelerate its convergence.

The early endeavors in the ISAC field focus on the co-existence of the communication system and the radar system \cite{saruthirathanaworakun2012opportunistic}. Nevertheless, the co-existence systems are inefficient in that signal processing is considered separately for communications and sensing. To remedy this defect, the co-existence system evolves into the dual-functional-radar-communication (DFRC) system \cite{ma2020joint,huang2020majorcom}. Various models have been proposed for the DFRC system in the past few years. One common model is the so-called collocated base station (BS) system, in which the communications function and the sensing function are performed at the same BS \cite{huang2020majorcom}. This model is further developed in  \cite{liuX2020joint} to account for the broadcast communications toward multiple receivers. For the DFRC model, \cite{liuX2020joint,he2023full,liu2021cramer} consider the point target sensing while \cite{wang2022partially,wang2024globally} consider the extended target sensing. Aside from the collocated BS case, the bi-static BS model has also been extensively studied in the literature to date \cite{guo2023bistatic1,zhu2023information,guo2023bistatic}. This model assumes that the signal transmission and the echo reception are conducted at two separate BSs; it has further developed into the notion of the multi-static BS system \cite{liu2024joint,li2023towards}.
Regarding the performance metrics, it is popular to use the Fisher information \cite{su2023sensing,wang2023stars}, and its reciprocal, i.e., the Cram{\'e}r-Rao bound (CRB) \cite{guo2023bistatic1,zhu2023information,guo2023bistatic}, to evaluate the performance of sensing. In contrast, the data rate is widely considered for communications \cite{guo2023bistatic1,zhu2023information}. Signal-to-noise ratio (SNR) is another common metric for sensing \cite{liu2021cramer,liuX2020joint,he2023full} and communications \cite{behdad2022power,liu2023snr,yu2023active}. This work focuses on maximizing a weighted sum of the data rates and the Fisher information in a multi-cell DFRC system with collocated BSs, but it can be readily adapted to other forms of optimization, e.g., maximizing the weighted sum rates under the CRB constraints.

The ISAC beamforming problem (particularly for the DFRC system) is a nontrivial task due to the nonconvex nature of the underlying optimization problem. Quite a few advanced optimization tools have been considered in the previous attempts. Semi-definite relaxation (SDR) is a typical example \cite{cheng2024optimal,liuX2020joint,liu2021cramer,yu2023active,zhou2023integrating,li2023integrated} because the ISAC beamforming problem can be somehow relaxed in a quadratic form, e.g., as a quadratic semi-definite program (QSDP) \cite{liuX2020joint}, a quadratically constrained quadratic program (QCQP) \cite{cheng2024optimal}, or a semi-definite program (SDP) \cite{liu2021cramer,yu2023active,zhou2023integrating,li2023integrated}. Successive convex approximation (SCA) constitutes another popular approach in this area \cite{huang2024edge,wang2022partially,meng2022throughput,he2023full}. For example, \cite{wang2022partially} uses SCA to convert the ISAC beamforming design to a second-order cone programming (SOCP) problem---which can be efficiently solved by the standard convex optimization method. Another line of studies \cite{hua2023secure,guo2023joint,chen2023joint} utilize the majorization-minimization (MM) theory to make the ISAC beamforming problem convex, especially when the passive beamforming of the intelligent reflecting surface (IRS) is involved. Moreover, because the ISAC beamforming problem is fractionally structured, the FP technique, i.e., the quadratic transform \cite{shen2018fractional1,shen2019optimization}, forms the building block of \cite{zou2024energy,deng2023beamforming}. Although \cite{guo2023bistatic,guo2023bistatic1} also treat the ISAC beamforming as an FP problem, their methods are based on the alternating direction method of multipliers method (ADMM) rather than the quadratic transform. Aside from the above model-based optimization approaches, deep learning has become a frontier technique for the ISAC beamforming, e.g., the long short-term memory network \cite{liu2022learning} and the unconventional convolutional neural network \cite{qi2024deep}.  

To exploit the degrees-of-freedom (DoF), the massive MIMO technology has been considered for the ISAC system. In principle, as shown in \cite{fortunati2020massive}, the efficiency of target detection improves with the antenna array size. From the algorithm design viewpoint, however, the large antenna array can pose a tough challenge. Actually, even considering the communications task alone, the beamforming design with massive antennas is already quite difficult. For instance, although the WMMSE algorithm \cite{cioffi_WMMSE,shi2011iteratively} has been extensively used for the MIMO transmission, it is no longer suited for the massive MIMO case because the algorithm then entails computing the large matrix inverse extensively per iteration. Other standard optimization methods such as SDR and SCA are faced with similar issues since they involve the matrix inverse operation implicitly when performing the interior-point optimization. Some recent efforts aim to improve the efficiency of WMMSE in the massive MIMO case, e.g., \cite{zhao2023rethinking} proposes a lite WMMSE algorithm that has a lower complexity of the matrix inverse computation under certain conditions. The present paper is most closely related to a series of recent works \cite{shen2023convergence,zhang2023enhancing} that use a nonhomogeneous bound from \cite{sun2016majorization} to avoid matrix inverse. While \cite{shen2023convergence,zhang2023enhancing} focus on the transmit beamforming alone, this work proposes a novel use of the nonhomogeneous bound that accounts for both the transmit beamforming and the echo receive beamforming.

The main novelties of this paper, especially as compared to \cite{shen2023convergence}, are summarized in the following:
\begin{itemize}
        \item Previous work \cite{shen2023convergence} considers the following problem:
        \begin{align*}
            \underset{x\in\mathcal X}{\text{maximize}}&\quad \sum_{i = 1}^{N}\log(1+\gamma_i(x)),
        \end{align*}
        where $\gamma_i(x)\ge0$ represents the SINR of link $i$. Notice that each SINR is a scalar for the single data stream case. In contrast, this work allows multi-stream transmission for each receiver, and thus each SINR term becomes a matrix; moreover, we need to deal with the traced matrix ratios of the Fisher information, so our problem is
        \begin{align*}
            \underset{x\in\mathcal X}{\text{maximize}}&\quad \sum_{i = 1}^{N}\log|\bm I+\bm\Gamma_i(x)|+\sum^L_{\ell=1}\beta_{\ell}\mathrm{tr}(\bm J_\ell(x)),
        \end{align*}
        where the weights $\beta_{\ell}\ge0$. In particular, when each $\beta_{\ell}=0$, the above problem reduces to the problem in \cite{zhang2023enhancing}.

        \item Both \cite{shen2023convergence} and \cite{zhang2023enhancing} assume that each BS has many antennas whereas each user terminal has a limited number of antennas. As a result, these two works just focus on eliminating the large matrix inverse for the transmitter side. In contrast, this work considers the case wherein all the devices have many antennas. As such, we propose a novel algorithm that eliminates the large matrix inverse for both transmitter and receiver sides. 
        
        \item We explore the connections between the proposed algorithm and the existing optimization methods. First, we show that the ISAC problem can be tackled by the conventional FP method---which has a connection with the WMMSE algorithm. Second, we show that the novel algorithm can be connected to the gradient projection method; the proof is based on a brand-new idea as shown in Section \ref{subsec:Fast FP}. As a result, the heavy-ball acceleration method in \cite{shen2023convergence} can be carried over from the log-scalar case to the log-det case.\\
\end{itemize}

\begin{table}[tbp]
\small
\renewcommand{\arraystretch}{1.4}
  \centering
  \caption{Main Variables Throughout the Paper}
    \begin{tabular}{ | m{3.2em} | m{18em}| } 
    \hline
    {\textbf{Symbol}} & {\textbf{Definition}} \\
    \hline
    \hline
     $L$ & Number of cells/BSs  \\
    \hline
    $K$      & Number of users in each cell \\
    \hline
    $N_\ell^t$      & Number of transmit antennas of BS $\ell$ \\
    \hline
    $N_\ell^r$      & Number of radar receive antennas of BS $\ell$ \\
    \hline
    $M_{\ell k}$      & Number of receive antennas of user $(\ell,k)$ \\
    \hline
    $\bfS_{\ell k}$      & Signal intended for user $(\ell,k)$ \\
    \hline
    $d_{\ell k}$      & Number of data streams for user $(\ell,k)$ \\
    \hline
    $T$      & Number of data frames \\
    \hline
    $\bfW_{\ell k}$      & Beamformer of BS $\ell$ for its $k$th user \\
    \hline
    $\bfH_{\ell k,i}$      & Channel from BS $i$ to user $(\ell,k)$ \\
    \hline
    {$\bfG_{\ell\ell}$} & {Response matrix of BS $\ell$}\\
     \hline
    {$\bfG_{\ell i}$} & {Interference channel from BS $i$ to BS $\ell$}\\
     \hline
    $\bm{\Delta}_{\ell k}$      &Background noise at user $(\ell,k)$\\
    \hline
    $\sigma^2$      & Background noise power at each user\\
    \hline
    {$\xi_{\ell}$}      & {Reflection coefficient with respect to BS $\ell$}\\
    \hline
    $\bm{a}_\ell^t(\theta_\ell)$   & Transmit steering vector of BS $\ell$ \\
    \hline
    $\bm{a}_\ell^r(\theta_\ell)$   & Receive steering vector of BS $\ell$\\
    \hline
     ${\bm \Psi}_{\ell k}$      &  Received signal of user $(\ell ,k)$\\
    \hline
    $\widetilde{\bm \Psi}_\ell$      & Received echo signal of BS $\ell$\\
    \hline
    $\widetilde{\bm{\Delta}}$      &Background noise at BS $\ell$ \\
    \hline
     $\tilde{\sigma}^2$      & Background noise power at BS $\ell$ \\
    \hline
    \end{tabular}%
  \label{tab:notation}%
\end{table}%

The remainder of this paper is organized as follows. Section \ref{sec:system model} introduces the system model and problem formulation. Section \ref{sec:joint beamforming} discusses the joint beamforming design. Section \ref{sec:simulation} presents numerical results. Finally, Section \ref{sec:conclusion} concludes the paper.
Here and throughout, bold lower-case letters represent vectors while bold upper-case letters represent matrices. For a vector $\bfa$ and  $\bfa^\bhh$ is its conjugate transpose.
For a matrix $\bfA$, $\bfA^*$ is its complex conjugate,  $\bfA^\top$ is its transpose, $\bfA^\text{H}$ is its conjugate transpose,  and $\|\bfA\|_F$ is its Frobenius norm. For a square matrix $\bfA$, $\mathrm{tr}(\bfA)$ is its trace, {$|\bfA|$ is its determinant}, and $\lambda_{\text{max}}(\bfA)$ is its largest eigenvalue. For a positive semi-definite matrix $\bfA$, $\sqrt{\bfA}$ is its square root, i.e., $\sqrt{\bfA}\sqrt{\bfA}^\bhh=\bfA$. Denote by $\bI_d$ the $d\times d$ identity matrix, $\mathbb C^{\ell}$ the set of $\ell\times1$ vectors, $\mathbb C^{d\times m}$ the set of $d\times m$ matrices,  $\mathbb H_{+}^{d\times d}$ the set of $d\times d$ positive semi-definite matrices, {and $\mathbb H_{++}^{d\times d}$ the set of $d\times d$ positive definite matrices}. For a complex number $a\in\mathbb C$, $\Re\{a\}$ is its real part, {$|a|$ is its absolute value.} We denote by $\otimes$ the Kronecker product, and $\mathrm{vec}(\cdot)$ the vectorization operation. The underlined letters represent the collections of the associated vectors or matrices, e.g., for $\bfa_1,\ldots,\bfa_n\in\mathbb C^d$ {we write $\underline\bfa = [\bfa_1, \bfa_2, 
\ldots, \bfa_n]^\top\in\mathbb C^{n\times d}$}. To ease reference, we list the main variables used in this paper in Table~\ref{tab:notation}.

\section{Multi-Cell ISAC System Model}
\label{sec:system model}

\begin{figure}[t]
\centering
\includegraphics[width=0.9\linewidth]{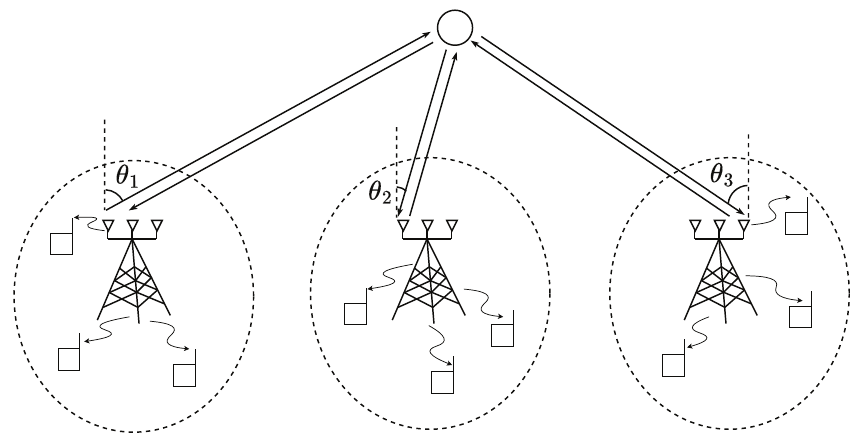}
\caption{A multi-cell ISAC system with $L=3$ and $K=3$. The circle is the point target to sense. The arrows are the transmit signals and the echo signals. Each BS $i$ aims to acquire $\theta_i$ independently.}
\label{fig:ISAC}
\end{figure}

Consider a downlink multi-cell ISAC system with $L$ cells. Assume that each cell has $K$ downlink users. For each BS $\ell=1,2,\ldots,L$, there are two tasks: (i) send independent messages to the $K$ downlink users in its cell by spatial multiplexing; (ii) detect the direction of arrival (DoA), $\theta_l$, for a point target. An illustrative example of the ISAC system as described above is shown in Fig.~\ref{fig:ISAC}.  We assume that BS $\ell$ has $N_\ell^t$ transmit antennas and $N^r_\ell$ receive antennas for detecting the echos. The $k$th downlink user in the $\ell$th cell is indexed as $(\ell,k)$. User $(\ell,k)$ has $M_{\ell k}$ receive antennas. Notably, the transmit antenna array size
$N_\ell^t$ and the echo antenna array size $N^r_\ell$ at the BS side can be large, whereas the receive antenna array size $M_{\ell k}$ at the terminal side is often small, as typically assumed for a massive MIMO network.

\subsection{Communications Model}
Let $\bfS_{\ell k}\in\mathbb C^{d_{\ell k}\times T}$ be the normalized symbol sequence intended for user $(\ell,k)$, where $d_{\ell k}$ is the number of data streams and $T$ is the block length. Note that $d_{\ell k}$ is often small since $d_{\ell k}\le M_{\ell k}$. Let $\bfW_{\ell k}\in\mathbb C^{N_\ell^t\times d_{\ell k}}$ be the transmit beamformer of BS $\ell$ for $\bfS_{\ell k}$.
The different data streams are assumed to be statistically independent, i.e.,
\begin{align}
	\frac{1}{T}\mathbb{E}\big[\bfS_{\ell k} \bfS_{\ell k}^\text{H}\big]  = \bI_{d_{\ell k}}.
\end{align}
Denote by $\bfH_{\ell k,i}\in \mathbb C^{M_{\ell k}\times N_i^t}$ the channel from BS $i$ to user $(\ell, k)$. It is assumed that channel state information, i.e., $\bfH_{\ell k,i}$ is perfectly known at the BSs\footnote{The robust beamforming case in the presence of channel estimation error is a much more complicated problem and deserves dedicated study, which we would like to consider in the future research.}. Each entry of the background noise $\bm{\Delta}_{\ell k}\in\mathbb{C}^{M_{\ell k}\times T}$ is drawn i.i.d. from $\mathcal{CN}(0,\sigma^2)$, and the received signal of user $(\ell,k)$ is given by
\begin{multline}
 \bm{\Psi}_{\ell k}=\bfH_{\ell k,\ell}\bfW_{\ell k}\bfS_{\ell k}
    +\underbrace{\sum_{j=1,j\ne k}^K\bfH_{\ell k,\ell}\bfW_{\ell j}\bfS_{\ell j}}_\text{intra-cell interference}
     \\+\underbrace{\sum_{i=1,i\ne \ell}^L\sum_{j= 1}^{K}\bfH_{\ell k,i}\bfW_{i j}\bfS_{i j}}_\text{cross-cell interference}
     +\bm{\Delta}_{\ell k}.
\end{multline} 
The data rate for user $(\ell,k)$ can be computed as\cite{goldsmith2005wireless}
\begin{align}
\label{eq:rate}
    R_{\ell k} = \log\Big|\bI_{d_{\ell k}}+\bfW_{\ell k}^\bhh\bfH_{\ell k,\ell}^\bhh\bfF_{\ell k}^{-1}\bfH_{\ell k,\ell}\bfW_{\ell k}\Big|,
\end{align}
where
\begin{multline}
\label{eq:com interference+noise}
    \bfF_{\ell k} = {\sum_{j=1,j\ne k}^K\bfH_{\ell k,\ell}\bfW_{\ell j}\bfW^\bhh_{\ell j}\bfH^\bhh_{\ell k,\ell}}\\
    +{\sum_{i=1,i\ne \ell}^L\sum_{j= 1}^{K}\bfH_{\ell k,i}\bfW_{i j}\bfW_{i j}^\bhh\bfH_{\ell k,i}^\bhh}+\sigma^2\bI_{M_{\ell k}}.
\end{multline}
For the good of communications, we wish to maximize the data rates throughout the network.

\subsection{Sensing Model}
Recall that BS $\ell$ has $N^t_\ell$ transmit antennas (for communications) and $N^r_\ell$ receive antennas (for sensing). Denote by $\bm{a}_{\ell}^t(\theta_{\ell})$ the steering vector of the transmit antennas, $\bm{a}_{\ell}^r(\theta_{\ell})$ the steering vector of the receive antennas, and $\xi_{\ell}$ the reflection coefficient for the path from  BS $\ell$ to the target then back to BS $\ell$. Define the response matrix for BS $\ell$ as
\begin{align}
    \bfG_{\ell  \ell}={\xi_{\ell}} \bm{a}_\ell^r(\theta_\ell)(\bm{a}_{ \ell}^t(\theta_{ \ell}))^\top,
\end{align}
where the transmit steering vector $\bm{a}_{\ell}^t(\theta_{\ell})$ and receive steering vector of $\bm{a}_\ell^r(\theta_\ell)$ can be specified as
\begin{align}
	\bm{a}_{\ell}^t(\theta_{\ell})=[1,e^{-j\pi\sin\theta_{\ell}},\ldots,e^{-j\pi(N_\ell^t-1)\sin\theta_{\ell}}]^\top,\\
	\bm{a}_{\ell}^r(\theta_{\ell})=[1,e^{-j\pi\sin\theta_{\ell}},\ldots,e^{-j\pi(N_\ell^r-1)\sin\theta_{\ell}}]^\top.
\end{align}
It is assumed that the exact DoA of the target is unknown at each BS. Rather, the BSs just ``roughly'' know the location of the target \cite{liu2021cramer,su2023sensing,hua2023mimo}, and aim to refine their knowledge through sensing. Denote by $\bG_{\ell i}$ the interference channel from BS $i$ to BS $\ell$ for any $i\ne \ell$. The echo signal $\widetilde{\bm{\Psi}}_\ell \in \mathbb{C}^{N_\ell^r \times T}$ received at BS $\ell$ is given by
\begin{align}
\widetilde{\bm{\Psi}}_\ell=\sum_{ i=1}^{L} \bG_{\ell i}\Bigg(\sum_{j=1}^{K}\bfW_{ij}\bfS_{ij}\Bigg)+\widetilde{\bm{\Delta}}_{\ell},
\end{align}
where each entry of the background noise $\widetilde{\bm{\Delta}}_{\ell}\in \mathbb C^{N_\ell^r \times T}$ is drawn i.i.d. from $\mathcal{CN}(0,\tilde{\sigma}^2)$. Only direct interference from other BSs is considered, the echo signals from other BSs are ignored due to their much lower power under the massive MIMO regime\cite{liu2020radar,liu2022optimal,wang2023qos}.

Mean squared error (MSE) is a common performance metric of estimation. Nevertheless, it is difficult to analyze the MSE of $\theta_\ell$ in our problem case. Instead, we adopt the Fisher information as the performance metric, following the previous works \cite{su2023sensing,wang2023stars} in the ISAC field. The target DoAs with respect to the different BSs are correlated, so the Fisher information ought to take this correlation into account if the different BSs collaborate to sense the target. Following the previous works\cite{shi2022device,liu2020joint,zhang2023joint}, we assume here that the different BSs perform the sensing tasks independently, i.e., each BS $\ell$  recovers the DoA from its own received echo signal $\widetilde{\bm\Psi}_{\ell}$.
Then the vectorization $\tilde{\bm{\psi}}_\ell=\mathrm{vec}({\widetilde{\bm{\Psi}}_\ell})$ can be computed as
\begin{align}
\label{eq:sensing signal}
	\tilde{\bm{\psi}}_\ell&=\sum_{k=1}^{K}(\bI_T \otimes \bfG_{\ell \ell}\bfW_{\ell k})\bfs_{\ell k}+\bm\mu_\ell,
\end{align}
where
\begin{align}
\bm\mu_\ell = \sum^L_{ i=1, i\neq \ell}\sum_{j=1}^{K}(\bI_T \otimes \bfG_{\ell  i}\bfW_{{ i}j})\bfs_{{ i}j}+\tilde{\bm{\delta}}_{\ell},
\end{align}
$\tilde{\bm{\delta}}_\ell=\mathrm{vec}({\widetilde{\bm{\Delta}}_{\ell}})$, 
and $\bfs_{\ell k} =\mathrm{vec}(\bfS_{\ell k})$. 
Define $\bfQ_{\ell}$ to be
\begin{align}
    \bm{Q}_\ell&= \mathbb E\big[\bm\mu_\ell\bm\mu_\ell^\text{H}\big]\notag\\
    &=\bI_T \otimes \Bigg(\sum^L_{i=1, i \neq \ell}\sum_{j=1}^{K}\bfG_{\ell  i}\bfW_{{ i}j}\bfW_{{ i}j}^\bhh\bfG_{\ell  i}^\bhh+ \tilde{\sigma}^2\bI_{N^r_{\ell}}\Bigg).\notag
\end{align}
Moreover, define $\bm{\upsilon}_{\ell}(\theta_{\ell}) = \sum_{k=1}^{K}(\bI_T \otimes \bfG_{\ell \ell}\bfW_{\ell k})\bfs_{\ell k}$. The Fisher information of $\theta_\ell$ is then given by\cite{van2002optimum}
\begin{align}
\label{eq:FI}
J_\ell&=2\left(\frac{\partial\bm{\upsilon}_{\ell}(\theta_{\ell})}{\partial\theta_{\ell}}\right)^\bhh\bfQ^{-1}_{\ell}\frac{\partial\bm{\upsilon}_{\ell}(\theta_{\ell})}{\partial\theta_{\ell}}\notag\\
&=2T \left\lbrace \sum_{k=1}^{K}\text{tr}\left((\dot{\bfG}_{\ell \ell}\bfW_{\ell k})^\text{H} \widehat\bfQ_\ell^{-1}( \dot{\bfG}_{\ell \ell}\bfW_{\ell k})\right)  \right\rbrace,
\end{align}
where  $\dot{\bfG}_{\ell \ell}=\partial \bfG_{\ell \ell}/\partial \theta_{\ell}$ and 
\begin{align}
\label{eq:Q hat}
   \widehat{\bm{Q}}_\ell= \sum_{ i=1,i\neq \ell}^L\sum_{j=1}^{K}\bfG_{\ell  i}\bfW_{{ i}j}\bfW_{{ i}j}^\bhh\bfG_{\ell  i}^\bhh+ \tilde{\sigma}^2\bI_{N^r_{\ell}}.
\end{align}
For the good of sensing, we wish to maximize the Fisher information of $\theta_\ell$ at each BS $\ell$, which is equivalent to minimizing the CRB on the MSE of $\theta_\ell$.

\subsection{ISAC Beamforming Problem}
To account for both communications and sensing, we consider maximizing a weighted sum of data rates and Fisher information:
\begin{subequations}
\label{prob:ISAC}
    \begin{align}
    \label{obj:ISAC}
        \underset{\underline{\bfW}}{\text{maximize}}\quad & \sum_{\ell=1}^L\sum_{k=1}^{K}\omega_{\ell k} R_{\ell k}+\sum_{\ell=1}^L\beta_{\ell}J_{\ell}\\
        \text{subject to}\quad & \sum_{k=1}^{K}\lVert \bfW_{\ell k}\rVert_F^2\le P_\ell,\;\forall \ell,
    \end{align}
\end{subequations}
where $\omega_{\ell k},\beta_\ell\ge 0$ are the given nonnegative weights reflecting the priorities of the communications and sensing tasks, and $P_\ell$ is the power budget of BS $\ell$.

We remark that adding up two distinct metrics (with weights) for the multiple-objective optimization purpose is fairly common in the literature. For instance, the optimization objective of the energy efficiency problem can be a weighted sum of data rates and power consumption \cite{shen2017flexible}, and the optimization objective of the caching problem can be a weighted sum of data rates and cache size \cite{tao2016content} or file delivery delay \cite{wu2019delay}. In the area of ISAC, the previous works \cite{su2023sensing,wang2024unified,gao2023cooperative} all consider a weighted sum of the communication performance metric and the sensing performance metric.

It is also quite common to formulate the ISAC problem as maximizing the communication performance metric under the constraint on the sensing performance metric, e.g., 
\begin{subequations}
    \begin{align}
        \underset{\underline{\bm W}}{\text{maximize}}&\quad \sum_{\ell = 1}^{L}\omega_{\ell k}R_{\ell k} \\
        \text{subject to}&\quad \,J_{\ell}\geq \Gamma_{\ell }\\
        &\quad \sum_{k=1}^K\|\bW_{\ell k}\|_F^2\leq P_{\ell},\;\forall \ell.
    \end{align}
\end{subequations}
We remark that the proposed FP method in Section \ref{sec:joint beamforming} still works for the above new problem. We can move the sensing performance constraint to the objective function by the Lagrange dual theory, then the existing ISAC beamforming problem can be solved immediately. Thus, the weight of the sensing performance metric in the multiple-objective optimization can be interpreted as a Lagrangian multiplier in the constrained optimization case. In practice, $\beta_\ell$ can be chosen empirically according to the QoS required for the sensing task.
\section{Main Results}
\label{sec:joint beamforming}

As a key observation, problem \eqref{prob:ISAC} is fractionally structured, i.e., each $R_{\ell k}$ contains an SINR inside the log-determinant while $J_\ell$ itself is a ratio, so the FP approach comes into play. Under this FP framework, we first develop a conventional FP approach to the ISAC beamforming problem. However, the complexity of the conventional FP becomes costly in the presence of the massive antenna array. To address this issue, we further propose a fast FP algorithm tailored to the ISAC beamforming with massive antennas.

\subsection{Preliminary}

We start with a quick review of the (matrix) FP technique, which is the building block of both the conventional FP algorithm and the proposed fast FP algorithm.
\setcounter{equation}{25}
\begin{figure*}
\vspace{-2em}
   \begin{align}
       \label{obj:ISAC:QT}
f_q(\underline{\bfW},\underline\bGa,\underline\bfY,\underline{\widetilde{\bfY}})=\sum_{\ell,k}\Big[\ttr\big(2\Re\{\bfW_{\ell k}^\bhh\bm{\Lambda}_{\ell k}\}-\omega_{\ell k}\bfY_{\ell k}^\bhh\bfU_{\ell k}\bfY_{\ell k}(\bI+\bGa_{\ell k}) -2T\beta_{\ell}\widetilde{\bfY}^{\bhh}_{\ell k}\widehat{\bfQ}_{\ell}\widetilde{\bfY}_{\ell k}\big)+\omega_{\ell k}\log|\bI+\bGa_{\ell k}|-\ttr(\omega_{\ell k}\bGa_{\ell k})\Big]
   \end{align}
   \hrule
\end{figure*}

For a pair of numerator function $\bfA_i(x)\in \mathbb H^{m\times m}_{+}$ and denominator function $\bfB_i(x)\in \mathbb H^{m\times m}_{++}$, the matrix ratio between them is
\setcounter{equation}{14}
\begin{align}
\label{matrix-ratio}
 \bfM_i(x) = \sqrt{\bfA}_i^{\bhh}(x)\bfB^{-1}_i(x)\sqrt{\bfA}_i(x),
\end{align} 
where $\sqrt{\bfA}_i(x)\in\mathbb{C}^{m\times d}$ is the square root of matrix $\bfA_i(x)$. We remark that $\bm H_{\ell k,\ell}\bm W_{\ell k}$ amounts to $\sqrt{\bA}_i(x)$, $\bm F_{\ell k}$ amounts to $\bB_i(x)$, $\dot{\bm G}_{\ell \ell}\bm W_{\ell k}$ amounts to $\sqrt{\bA}_i(x)$, and $\widehat{\bm Q}_{\ell}$ amounts to $\bB_i(x)$. We now consider $n$ matrix ratios and formulate the \emph{sum-of-weighted-log-ratios} problem as 
 \begin{align}
\underset{x\in\mathcal X}{\text{maximize}} &\quad\sum^n_{i=1} \omega_i\log\left|\bI_d+\bfM_{i}(x)\right|
\label{prob:log ratio}
\end{align}   
with the positive weights $\omega_{i}>0$ and a nonempty constraint set $\mathcal{X}$. It turns out that the matrix ratios can be ``moved'' to the outside of logarithms.
\begin{proposition}[Lagrangian Dual Transform \cite{shen2019optimization}]
\label{prop:LDT}
Problem \eqref{prob:log ratio}
can be recast to
\begin{subequations}
\begin{align}
\underset{x\in\mathcal X,\,\underline\bGa}{\text{maximize}} &\quad f_r(x,\underline{\bGa})\\
  \text{subject to} &\quad  \bGa_i\in\mathbb {H}^{m\times m}_+,
\end{align}
\end{subequations}
where
\begin{align}
f_r(x,\underline{\bGa})=&\sum_{i=1}^n\omega_i\Big(\log|\bI_d+\bGa_i|-\ttr\big(\bGa_i\big)+\ttr\Big((\bI_d+\bGa_i)\cdot\notag\\
&\sqrt{\bfA}_i^{\bhh}(x)\big(\bfA_i(x)+\bfB_i(x)\big)^{-1}\sqrt{\bfA}_i^{\bhh}(x)\Big)\Big).    
\end{align}
\end{proposition}
Moreover, given a sequence of matrix coefficients
$\bfC_i\in\mathbb{H}^{m\times m}_{++}$, let us consider the \emph{sum-of-weighted-traces-of-matrix-ratio} problem:
\setcounter{equation}{18}
\begin{align}
\label{prob:sum-weighted-ratios}
    \underset{x\in\mathcal{X}}{\text{maximize}}&\quad\sum_{i=1}^n\omega_i\ttr(\bfC_i\bfM_i(x)).
\end{align}
We now decouple the matrix ratios in the above problem.
\begin{proposition}[Quadratic Transform \cite{shen2019optimization}]
    \label{prop:QT}
    Problem \eqref{prob:sum-weighted-ratios} can be recast to 
\begin{subequations}
\begin{align}
\underset{x\in\mathcal X,\,\underline\bfY}{\text{maximize}} &\quad f_q(x,\underline{\bfY})\\
  \text{subject to} &\quad \bY_i\in\mathbb{C}^{m \times d},
\end{align}
\end{subequations}
where
\begin{align}
&f_q(x,\underline{\bfY})
    =\notag\\
    &\quad\sum_{i=1}^n\omega_i\ttr\left(2\Re\{\sqrt{\bfA}^\bhh_i(x)\bfY_i\bfC_i\}-\bfY_i^\bhh\bfB_i(x)\bfY_i\bfC_i\right).
\end{align}
\end{proposition}
We remark that the above proposition slightly extends the original form of the quadratic transform in \cite{shen2019optimization} by incorporating the matrix coefficients $\bfC_i$. Notice that it is critical to assume $\bfC_i\succ\bm0$; the optimal $\bY_i$ may not be unique if $\bfC_i\succeq\bm0$.

\subsection{Conventional FP for ISAC Beamforming}
\label{subsec:FP}

We now return to the ISAC beamforming problem in \eqref{prob:ISAC}. First, we use the Lagrangian dual transform in Proposition \ref{prop:LDT} to move the ratios to the outside of log-determinants for the term $\sum_{\ell=1}^L\sum_{k=1}^{K}\omega_{\ell k} R_{\ell k}$ in \eqref{obj:ISAC}. As a result, problem \eqref{prob:ISAC} is converted to
\setcounter{equation}{21}
\begin{subequations}
\label{prob:ISAC:LDT}
   \begin{align}
   \label{obj:ISAC:LDT}
      \underset{\underline{\bfW},\,\underline{\bGa}}{\text{maximize}}\quad&f_r(\underline\bfW,\underline{\bGa})+\sum_{\ell=1}^L\beta_{\ell}J_{\ell}\\
    \text{subject to}\quad & \sum_{k=1}^{K}\lVert \bfW_{\ell k}\rVert_F^2\le P_\ell,\;\forall \ell\\
    & \,\bGa_{\ell k}\in\mathbb H^{d_{\ell k}\times d_{\ell k}}_{+},
\end{align} 
\end{subequations}
where
\begin{multline}
\label{obj:rate:LDT}
     f_r(\underline\bfW,\underline\bGa) = \sum^L_{\ell=1}\sum^K_{k=1}\omega_{\ell k}\bigg[\log\big|\bI_{d_{\ell k}}+\bGa_{\ell k}\big|-\ttr\big(\bGa_{\ell k}\big)\\+\ttr\Big((\bI_{d_{\ell k}}+\bGa_{\ell k})\bfW_{\ell k}^\bhh\bfH_{\ell k,\ell}^\bhh\bfU_{\ell k}^{-1}\bfH_{\ell k,\ell}\bfW_{\ell k}\Big)\bigg]
\end{multline}
and
\begin{align}
    \bfU_{\ell k} = \sum_{i=1}^L\sum_{j=1}^K\bfH_{\ell k,i}\bfW_{ij}\bfW^\bhh_{ij}\bfH^\bhh_{\ell k,i}+\sigma^2\bI_{M_{\ell k}}.
\end{align}
When $\underline\bfW$ is fixed, problem \eqref{prob:ISAC:LDT} is convex in $\underline\bGa$. According to first-order condition, each $\bGa_{\ell k}$ can be optimally determined as
\begin{align}
\label{eq:update_Ga}
    \bGa_{\ell k}^\star = \bfW_{\ell k}^\bhh\bfH_{\ell k,\ell}^\bhh\bfF_{\ell k}^{-1}\bfH_{\ell k,\ell}\bfW_{\ell k}.
\end{align}
We now consider optimizing $\underline\bfW$ for fixed $\underline\bGa$. 

Notice that \eqref{prob:ISAC:LDT} is a sum-of-weighted-traces-of-matrix-ratio problem of $\underline\bfW$ when $\underline\bGa$ is fixed, so the quadratic transform from Proposition \ref{prop:QT} is applicable. The optimization objective in \eqref{obj:ISAC:LDT} is then further recast to $f_q(\underline{\bfW},\underline\bGa,\underline\bfY,\underline{\widetilde{\bfY}})$ as shown in \eqref{obj:ISAC:QT} with
\setcounter{equation}{26}
\begin{align}
\label{matrix:Lambda}
    \bm\Lambda_{\ell k} = {\omega}_{\ell k}\bfH_{\ell k,\ell}^\bhh\bfY_{\ell k}(\bI_{d_{\ell k}}+\bGa_{\ell k})+2T\beta_\ell\dot{\bfG}_{\ell \ell}^\text{H}\widetilde{\bfY}_{\ell k}.
\end{align}
For the notational clarity in \eqref{obj:ISAC:QT}, we use $\bY_{\ell k}$ to denote each auxiliary variable introduced (by the quadratic transform) for the matrix ratios related to the communication task, and use $\widetilde{\bY}_{\ell k}$ to denote each auxiliary variable introduced for the matrix ratios related to the sensing task. The resulting further reformulation of problem  \eqref{prob:ISAC:LDT} is
\setcounter{equation}{27}
\begin{subequations}
\label{prob:ISAC:QT}
    \begin{align}
\underset{\substack{\underline{\bfW},\,\underline\bGa,\,\underline\bfY,\,\underline{\widetilde{\bfY}}}}{\text{maximize}}\quad&f_q(\underline{\bfW},\underline\bGa,\underline\bfY,\underline{\widetilde{\bfY}})\\
\text{subject to}\quad & \sum_{k=1}^{K}\lVert \bfW_{\ell k}\rVert_F^2\le P_\ell,\;\forall \ell\\
&\, \bGa_{\ell k}\in\mathbb H^{d_{\ell k}\times d_{\ell k}}_{+}\\
&\, \bY_{\ell k}\in\mathbb{C}^{M_{\ell k}\times d_{\ell k}}\\ 
&\,\widetilde\bY_{\ell k}\in\mathbb{C}^{N_{\ell}^r\times d_{\ell k}}.
\end{align}
\end{subequations}
When $\underline\bfW$ and $\underline\bGa$ are both held fixed, the above problem is jointly convex in $\underline\bfY$ and $\underline{\widetilde\bfY}$, so we can optimally determine them by the first-order condition as
\begin{align}
  \bfY^\star_{\ell k}&=\bfU_{\ell k}^{-1}{\bfH_{\ell k,\ell}\bfW_{\ell k}},
    \label{eq:update_Y}\\
    \widetilde{\bfY}^\star_{\ell k}&=\widehat{\bfQ}_\ell^{-1}\dot{\bfG}_{\ell \ell}\bfW_{\ell k}.
    \label{eq:update_tilde_Y}   
\end{align}
We then aim to optimize $\underline\bfW$ with the rest variables held fixed. Toward this end, we rewrite $f_q(\underline{\bfW},\underline\bGa,\underline\bfY,\underline{\widetilde{\bfY}})$ as
\begin{multline}
\label{obj:ISAC:QT:re}
f_q(\underline{\bfW},\underline\bGa,\underline\bfY,\underline{\widetilde{\bfY}})=\sum_{\ell,k}\ttr\left(2\Re\{\bfW_{\ell k}^\bhh\bm{\Lambda}_{\ell k}\} -\bfW_{\ell k}^\bhh\bfL_\ell\bfW_{\ell k}\right)\\+\mathrm{const},
\end{multline}
where $\mathrm{const}$ is a constant term when $(\underline\bGa,\underline\bfY,\underline{\widetilde{\bfY}})$ are fixed, and
\begin{multline}
\label{matrix L}
    \bfL_{\ell} = \sum_{i=1}^L\sum_{j=1}^{K}\omega_{i j}\bfH_{i j,\ell}^\bhh\bfY_{ij}(\bI_{d_{ij}}+\bGa_{ij})\bfY_{i j}^\bhh\bfH_{i j,\ell}\\+2T\sum^L_{i=1,i\ne \ell}\sum_{j=1}^{K}\bfG_{i \ell}^\bhh(\beta_{i}\widetilde{\bfY}_{ij}\widetilde{\bfY}_{ij}^\bhh)\bfG_{i \ell}.
\end{multline}
By the identity $\ttr(\bfA\bfA^\bhh)=\|\bfA\|_F^2$, we get the optimal $\bfW_{\ell k}$:
\begin{align}
\label{eq:WMMSE solve W}
    \bfW_{\ell k}^\star = \arg\underset{\underline{\bfW}\in\mathcal{W}}{\min}\|\bfL_{\ell}^{\frac{1}{2}}(\bfW_{\ell k}-\bfL_{\ell}^{-1}\bm{\Lambda}_{\ell k})\|^2_F,
\end{align}
where 
\begin{align}
\label{set W}
    \mathcal{W}=\left\{\underline{\bfW}:\sum_{k=1}^{K}\lVert \bfW_{\ell k}\rVert_F^2\le P_\ell,\;\forall \ell\right\}
\end{align}
is the feasible set.
The above solution can be further written as
\begin{align}
\label{eq:update_W}
    \bfW^\star_{\ell k} = \left(\eta_{\ell}\bI_{N_\ell^t}+\bfL_{\ell}\right)^{-1}\bm\Lambda_{\ell k},
\end{align}
where the Lagrange multiplier $\eta_\ell$ accounts for the power constraint and can be optimally determined as
\begin{align}
\label{eq:update_eta}
    \eta_{\ell} = \min\Bigg\{\eta \ge 0:\sum_{{k=1}}^{K}\|\bfW_{\ell k}(\eta)\|_F^2\le P_{\ell}\Bigg\}.
\end{align}
We summarize the above iterative optimization steps in Algorithm \ref{algorithm:WMMSE}, which we refer to as the conventional FP method for the ISAC beamforming. The following three remarks in order discuss the connection between Algorithm \ref{algorithm:WMMSE} and the WMMSE algorithm \cite{shi2011iteratively},  the bottleneck of Algorithm \ref{algorithm:WMMSE}, and how the proposed method differs from the existing FP-based approaches to the ISAC \cite{guo2023bistatic1}, \cite{guo2023bistatic}. 

\begin{remark}
    If each $\beta_\ell=0$, i.e., when we consider maximizing the weighted sum rates alone, then Algorithm \ref{algorithm:WMMSE} reduces to the WMMSE algorithm \cite{cioffi_WMMSE,shi2011iteratively}. Nevertheless, the WMMSE was initially proposed in \cite{cioffi_WMMSE} based on a duality between the rate maximization and the MSE minimization, whereas we rederive it by the FP technique.
\end{remark}

\begin{remark}
    Notice that updating $\underline\bfW$ as in \eqref{eq:update_W} and updating $\underline{\widetilde\bY}$ as in \eqref{eq:update_tilde_Y} all entail computing the matrix inverses, which can be quite costly when massive antennas are deployed so that $N^r_{\ell}$ or/and $N^t_{\ell}$ is large. The nonhomogeneous FP algorithm proposed in Section \ref{nonhomogeneous ISAC} aims to eliminate the matrix inverse operation.
\end{remark}
\begin{remark}
    The previous works \cite{guo2023bistatic1}, \cite{guo2023bistatic} and our work both consider the ISAC problem from an FP perspective. Nevertheless, their FP problem types are fundamentally different. In principle, \cite{guo2023bistatic1}, \cite{guo2023bistatic} consider the scalar-ratio problems,
whereas our work considers the matrix-ratio problem.
This is because our model allows each user terminal to have multiple receive antennas to support MIMO transmission, whereas \cite{guo2023bistatic1}, \cite{guo2023bistatic} assume that each user terminal has a single receive antenna. Moreover, the sensing models are different. The previous works \cite{guo2023bistatic1}, \cite{guo2023bistatic} assume a bi-static sensing system with $\xi$ unknown, whereas this work follows \cite{cheng2019co,chen_TSP,cheng2022qos} assuming a co-located sensing system with $\xi$ available. 
\end{remark}
\begin{algorithm}[t]
  \caption{Conventional FP for ISAC Beamforming}
  \label{algorithm:WMMSE}
  \begin{algorithmic}[1]
      \STATE Initialize $\underline{\bfW}$ to a feasible value.
      \REPEAT 
      \STATE Update each $\bm{\Gamma}_{\ell k}$ by \eqref{eq:update_Ga}.
      \STATE Update each $\bfY_{\ell k}$ and $\widetilde{\bfY}_{\ell k}$ by \eqref{eq:update_Y} and \eqref{eq:update_tilde_Y}, respectively.
      \STATE Update each $\bfW_{\ell k}$ by \eqref{eq:update_W}.
      \UNTIL{the objective value converges} 
  \end{algorithmic}
\end{algorithm}

\subsection{Large Matrix Inverse Elimination}
\label{subsec:FFP}

Before tackling the large matrix inverse issue of the ISAC beamforming problem, we first illustrate how this issue can be addressed in a toy example. The main tool is stated in the following lemma:
\begin{lemma}[Nonhomogeneous Bound]
\label{lemma:Nonhomo}
 Suppose that the two Hermitian matrices $\bfL,\bfK\in\mathbb C^{d\times d}$ satisfy $\bfL\preceq\bfK$, e.g., when $\bfK=\lambda\bI$ where $\lambda\ge\lambda_{\text{max}}(\bfL)$. Then for any two matrices $\bfX,\bfZ\in\mathbb C^{d\times m}$, one has
\begin{multline}
\label{Taylor_inequality}
\ttr(\bfX^\bhh\bfL\bfX)\le
\ttr\big(\bfX^\bhh\bfK\bfX+2\Re\{\bfX^\bhh(\bfL-\bfK)\bfZ\}\\+\bfZ^\bhh(\bfK-\bfL)\bfZ\big),
\end{multline}
where the equality holds if $\bfZ=\bfX$.
\end{lemma}
\begin{IEEEproof}
Because $f(\bfX)=\ttr(\bfX^\bhh(\bfK-\bfL)\bfX)$ is convex, we have
$$
f(\bfX) \ge f(\bfZ) + \Re\left\{\ttr\left[(\nabla f(\bfZ))^\top(\bfX-\bfZ)\right]\right\}, 
$$
where $\nabla f(\bfZ)=2(\bfK-\bfL)^\top\bfZ^*$, and thereby have \eqref{Taylor_inequality}.
\end{IEEEproof}

Notice that the above lemma generalizes the result of Example 13 in \cite{sun2016majorization}. 
\begin{remark}
\label{remark:lambda}
    One option is to let $\lambda=\lambda_{\max}$ exactly. The value of $\lambda_{\max}$ can be efficiently obtained by the power method \cite{datta2004numerical}. However, there can be a numerical precision issue so that $\bfL\preceq\bfK$ may not hold in practice. To remedy this situation, we can consider using an upper bound on $\lambda_{\max}$, e.g., $\mathrm{tr}(\bfL)$ and $\|\bfL\|_F$. In particular, assuming $\bm L\succeq \bm 0$, we have
$$\mathrm{tr}(\bm L)\geq \|\bfL\|_F\geq\lambda_{\max}(\bm L),$$
so the Frobenius norm is a tighter upper bound. Nevertheless, the computational complexity of the Frobenius norm is higher than that of the trace bound. Thus, there is a tradeoff between efficiency and precision. In practice, how $\lambda_{\max}$ is approximated or evaluated can be a case-by-case decision, depending on the specific requirements about the per-iteration complexity and the number of iterations.
\end{remark}

The remainder of Section \ref{subsec:FFP} shows how Lemma \ref{lemma:Nonhomo} can be used to eliminate the matrix inverse from the iterative optimization of a single-ratio problem. Then Section \ref{nonhomogeneous ISAC} extends this result to the multi-ratio case in the ISAC beamforming problem.

\begin{example}
\label{example:1}
Consider the following single-ratio problem:
    \begin{align}
    \label{prob:example fo}
\underset{\bfX\in\mathcal{X}}{\text{maximize}}\quad&\ttr((\bfA\bfX)^\bhh(\bfB\bfX\bfX^\bhh\bfB^\bhh)^{-1}(\bfA\bfX)),
\end{align}
where $\bfA\in\mathbb C^{n\times d}$, $\bfB\in\mathbb C^{n\times d}$, $\bfX\in\mathbb C^{d\times m}$, and $\mathcal X$ is a nonempty constraint set on $\bfX$. By the quadratic transform in Proposition \ref{prop:QT}, the above problem can be recast to
\begin{subequations}
\label{prob:example fq}
    \begin{align}
\underset{\bfX\in\mathcal X,\bfY}{\text{maximize}}\quad&f_q(\bfX,\bfY)\\
\text{subject to}\quad & \bfY\in\mathbb C^{n\times m},
\end{align}
\end{subequations}
where the new objective function is
\begin{align}
f_q(\bfX,\bfY) = \ttr\big(2\Re\{\bfX^\bhh(\bfA^\bhh\bfY)\}-\bfX^\bhh(\bfB^\bhh\bfY\bfY^\bhh\bfB)\bfX\big).
\end{align}
If we optimize $\bfX$ and $\bfY$ iteratively in the above new problem, then both $\bfX$ and $\bfY$ can be optimally updated in closed form by completing the square in $f_q(\bfX,\bfY)$. However, these optimal updates incur the matrix inverse operation, which can be computationally costly when $n$ and $d$ are large. The classic WMMSE algorithm is faced with the same issue.

\setcounter{equation}{52}
\begin{figure*}
    \begin{align}
      \label{ISAC g_o}
g_o(\underline{\bfW},\underline\bGa,\underline\bfY,\underline{\widetilde\bfY},\underline{\bfZ})=\sum_{\ell,k}\big[\ttr\left(2\Re\{\bfW_{\ell k}^\bhh\mathbf{\Lambda}_{\ell k}+\bfW_{\ell k}^\bhh(\lambda_{\ell}\bI_{N_{\ell}^t}-\bfL_{\ell})\bfZ_{\ell k}\}+\bfZ_{\ell k}^\bhh(\bfL_{\ell}-\lambda_{\ell}\bI_{N_{\ell}^t})\bfZ_{\ell k} -\lambda_{\ell}\bfW_{\ell k}^\bhh\bfW_{\ell k}\right)
    \notag\\ -\ttr\left(2\tilde{\sigma}^2T\beta_\ell\widetilde\bfY_{\ell k}^\text{H}\widetilde\bfY_{\ell k}+\omega_{\ell k}\sigma^2(\bI_{d_{\ell k}}+\bGa_{\ell k})\bfY_{\ell k}^\bhh\bfY_{\ell k}\right)+\omega_{\ell k}\log|\bI_{d_{\ell k}}+\bGa_{\ell k}|-\ttr(\omega_{\ell k}\bGa_{\ell k})\big]
    \end{align}
    \hrule
\end{figure*}
\setcounter{equation}{56}
\begin{figure*}
\vspace{-2em}
\begin{multline}
   \label{obj:ISAC:GQT}
g_s(\underline{\bfW},\underline\bGa,\underline\bfY,\underline{\widetilde{\bfY}},\underline{\bfZ},\underline{\widetilde\bfZ})=\sum_{\ell,k}\bigg[\ttr\big(\Re\{2\bfW_{\ell k}^\bhh\bm{\Lambda}_{\ell k}+2\widetilde\bfY_{\ell k}^\bhh(\tilde\lambda_{\ell}\bI_{N_{\ell}^r}-\widetilde\bfL_{\ell})\widetilde\bfZ_{\ell k}+\widetilde\bfZ_{\ell k}^\bhh(\widetilde\bfL_{\ell}-\tilde\lambda_{\ell}\bI_{N_{\ell}^r})\widetilde\bfZ_{\ell k}-(2\bfW_{\ell k}-\bfZ_{\ell k})^\bhh\bfD_{\ell}\bfZ_{\ell k}
\\
+\lambda_{\ell}(2\bfW_{\ell k}^\bhh\bfZ_{\ell k}-\bfZ_{\ell k}^\bhh\bfZ_{\ell k}-\bfW_{\ell k}^\bhh\bfW_{\ell k})\}\big) +\omega_{\ell k}\log|\bI_{d_{\ell k}}+\bGa_{\ell k}|-\ttr\big(\omega_{\ell k}\bGa_{\ell k}+\omega_{\ell k}\sigma^2(\bI_{d_{\ell k}}+\bGa_{\ell k})\bfY_{\ell k}^\bhh\bfY_{\ell k}\big)\bigg]
\end{multline}
\hrule
\end{figure*}
\setcounter{equation}{40}

Now we propose reformulating problem \eqref{prob:example fq} further by the nonhomogeneous bound in Lemma \ref{lemma:Nonhomo}. Treating 
\begin{equation}
  \bfL = \bfB^\bhh\bfY\bfY^\bhh\bfB,
\end{equation}
we can convert problem \eqref{prob:example fq} to 
\begin{subequations}
\label{prob:example go}
    \begin{align}
\underset{\bfX\in\mathcal X,\bfY,\bfZ}{\text{maximize}}\quad&g_o(\bfX,\bfY,{\bfZ})\\
\text{subject to}\quad & \bfY\in\mathbb C^{n\times m}\\
& {\bfZ}\in\mathbb C^{d\times m},
\end{align}
\end{subequations}
where the new objective function is
\begin{multline}
\label{example g_o}
    g_o(\bfX,\bfY,{\bfZ})=\ttr\Big(2\Re\{\bfX^\bhh(\bfA^\bhh\bfY)+\bfX^\bhh(\lambda\bI_d-\bfL)\bfZ\}\\
+\bfZ^\bhh(\bfL-\lambda\bI_d)\bfZ-\lambda\bfX^\bhh\bfX\Big)
\end{multline}
with $\lambda=\lambda_{\text{max}}(\bfL)$. Notice that $\bfX$ can now be optimally determined for $g_o(\bfX,\bfY,{\bfZ})$ in closed form without computing the matrix inverse when the other variables $(\bfY,{\bfZ})$ are held fixed. Subsequently, $\bfZ$ can be determined in closed form according to Lemma 1 as $\bfZ^\star = \bfX.$ This desirable result motivates us to apply Lemma \ref{lemma:Nonhomo} one more time in order to get rid of the matrix inverse for the optimal update of $\bfY$ in solving \eqref{prob:example go}.  Specifically, rewriting \eqref{example g_o} as
\begin{multline}
\label{example g_o re}
g_o(\bfX,\bfY,{\bfZ}^\star)=\ttr\Big(\Re\{-\bfY^\bhh(\bfB\bfZ^\star(2\bfX-{\bfZ}^\star)^\bhh\bfB^\bhh)\bfY\\
+2\bfY^\bhh(\bfA\bfX)+\lambda(2\bfX^\bhh\bfZ^\star-\bfX^\bhh\bfX-(\bfZ^\star)^\bhh\bfZ^\star)\}\Big)
\end{multline}
and treating
\begin{equation}
    \widetilde\bfL=\bfB\bfZ(2\bfX-{\bfZ}^\star)^\bhh\bfB^\bhh,
\end{equation}
we further recast problem \eqref{prob:example go} to
\begin{subequations}
\label{prob:example g_s}
    \begin{align}
\underset{\bfX\in\mathcal X,\bfY,\bfZ^\star,\widetilde{\bfZ}}{\text{maximize}}\quad&g_s(\bfX,\bfY,\bfZ^\star,\widetilde{\bfZ})\\
\text{subject to}\quad & \bfY\in\mathbb C^{n\times m}\\
& \bfZ^\star =\bfX\\
& \widetilde{\bfZ}\in\mathbb C^{n\times m},
\end{align}
\end{subequations}
where 
\begin{multline}
\label{example g_s}
g_s(\bfX,\bfY,{\bfZ}^\star,\widetilde{\bfZ})= \ttr\big(\Re\{2\bfY^\bhh(\bfA\bfX+(\tilde\lambda\bI_n-\widetilde\bfL){\widetilde\bfZ})\\
+\widetilde\bfZ^\bhh(\widetilde\bfL-{\tilde\lambda}\bI_n)\widetilde\bfZ+\lambda(2\bfX^\bhh\bfZ^\star-\bfX^\bhh\bfX-(\bfZ^\star)^\bhh\bfZ^\star)\\
-\tilde\lambda\bfY\bfY^\bhh\}\big)
\end{multline}
with $\tilde\lambda=\lambda_{\text{max}}(\widetilde\bfL)$. We propose optimizing the variables of $g_s(\bfX,\bfY,\bfZ,\widetilde{\bfZ})$ in an iterative fashion as
\begin{align*}
    \cdots\rightarrow\bfX^\tau \rightarrow \bfZ^\tau\rightarrow\bfY^\tau \rightarrow{\widetilde\bfZ}^\tau\rightarrow \bfX^{\tau+1} \rightarrow \cdots.
\end{align*} 
We now specify the iterative optimization steps. First, according to Lemma \ref{lemma:Nonhomo}, $\bfZ$ and $\widetilde{\bfZ}$ are optimally updated as
\begin{align}
\bfZ^\star&=\bfX,\\
\widetilde\bfZ^\star&=\bfY.
\end{align}
With the optimal $\widetilde\bfZ^\star=\bfY$ plugged in $g_s(\bfX,\bfY,\widetilde{\bfZ},\bfZ)$, we can find the optimal update of $\bfX$ by completing the square as
\begin{align}
\bfX^\star&=\arg\underset{{\bfX}\in\mathcal{X}}{\min}\|\lambda\bfX-\bfA^\bhh\bfY-(\lambda\bI_d-\bfL)\bfZ\|_F^2\notag\\
&=\mathcal{P}_{\mathcal{X}}\big(\bfZ+\frac{1}{\lambda}(\bfA^\bhh\bfY-\bfL\bfZ)\big).
\end{align}
Likewise, after $\bfZ$ has been optimally updated to $\bfX$, we can find the optimal update of $\bfY$ by completing the square as
\begin{align}
\bfY^\star&=\arg\min\|\tilde\lambda\bfY-\bfA\bfX-(\tilde\lambda\bI_n-\widetilde\bL)\widetilde\bfZ\|_F^2\notag\\
&=\widetilde\bfZ+\frac{1}{\tilde\lambda}(\bfA\bfX-\widetilde\bfL\widetilde\bfZ).
\end{align}
We remark that the above iterative optimization steps do not incur any matrix inverse.
\end{example}

\subsection{Nonhomogeneous FP for ISAC Beamforming}
\label{nonhomogeneous ISAC}

We now extend the result of Example \ref{example:1} to the multi-ratio FP case for the ISAC beamforming. First, we still apply the Lagrangian dual transform in Proposition \ref{prop:LDT} to problem \eqref{prob:ISAC} as formerly shown in Section \ref{subsec:FP}, so that the original problem is converted to \eqref{prob:ISAC:LDT}. With $\underline\bGa$ iteratively updated as in \eqref{eq:update_Ga}, solving for $\underline\bfW$ in \eqref{prob:ISAC:LDT} boils down to a sum-of-ratios problem.

Again, we then decouple each ratio by the quadratic transform in Proposition \ref{prop:QT}. The resulting further reformulated problem is formerly shown in \eqref{prob:ISAC:QT}. Recall that if we stop here and consider optimizing the primal variable $\underline\bfW$ along with the auxiliary variables $(\underline\bGa,\underline\bfY,\underline{\widetilde\bfY})$ in an iterative fashion, then we end up with a conventional FP algorithm---which incurs large matrix inverse.

In order to get rid of the large matrix inversion, we follow the procedure in Example \ref{example:1} and reformulate problem \eqref{prob:ISAC:QT} further. Treating each $\bfL_{\ell}$ in \eqref{matrix L} as $\bfL$ in \eqref{Taylor_inequality}, we can use Lemma \ref{lemma:Nonhomo} to reformulate problem \eqref{prob:ISAC:QT} as
\setcounter{equation}{51}
\begin{subequations}
\label{prob:ISAC go}
    \begin{align}
\underset{\underline\bfW,\underline\bGa,\underline\bfY,\underline{\widetilde\bfY},\underline{{\bfZ}}}{\text{maximize}}\quad&g_o(\underline\bfW,\underline\bGa,\underline\bfY,\underline{\widetilde\bfY},\underline{{\bfZ}})\\
\text{subject to}\quad & \underline\bfW\in\mathcal W \label{cons:W}\\
& \bGa_{\ell k}\in\mathbb H^{d_{\ell k}\times d_{\ell k}}_{+}\\
& \bY_{\ell k}\in\mathbb{C}^{M_{\ell k}\times d_{\ell k}}\\ 
&\widetilde\bY_{\ell k}\in\mathbb{C}^{N_{\ell}^r\times d_{\ell k}}\\
&{\bfZ}_{\ell k}\in\mathbb {C}^{N_{\ell}^t\times d_{\ell k}},\label{cons:Z}
\end{align}
\end{subequations}
where $\mathcal W$ is previously defined in \eqref{set W}, and the new objective function $g_o(\underline\bfW,\underline\bGa,\underline\bfY,\underline{\widetilde\bfY},{\underline{\bfZ}})$ is shown in \eqref{ISAC g_o} as displayed at the top of the page with $\lambda_{\ell}=\lambda_{\text{max}}(\bfL_{\ell})$. Note that the above $g_o(\underline\bfW,\underline\bGa,\underline\bfY,\underline{\widetilde\bfY},\underline{{\bfZ}})$ amounts to a multi-ratio generalization of the previous $g_o(\bfX,\bfY,{\bfZ})$ in \eqref{example g_o} in Example \ref{example:1}.

Next, defining 
\setcounter{equation}{53}
\begin{align}
    \bfD_{\ell} = \sum_{i=1}^L\sum_{j=1}^{K}\omega_{\ell j}\bfH_{i j,\ell}^\bhh\bfY_{ij}(\bI_{d_{ij}}+\bGa_{ij})\bfY_{i j}^\bhh\bfH_{i j,\ell}, 
\end{align}
and treating 
\begin{multline}
\widetilde\bfL_{\ell} = T\beta_{\ell}\Big(\sum^L_{i=1,i\ne \ell}\sum_{j=1}^{K}\bfG_{\ell i}({\bfZ}_{ij}(2{\bfW}_{ij}-\bfZ_{ij})^\bhh)\bfG^\bhh_{\ell i}+ \tilde{\sigma}^2\bI\Big)
\end{multline}
for each $\ell$ as $\bfL$ in \eqref{Taylor_inequality}, we use Lemma \ref{lemma:Nonhomo} to further reformulate \eqref{prob:ISAC go} as
\begin{subequations}
\label{prob:ISAC gs}
    \begin{align}
\underset{\underline\bfW,\underline\bGa,\underline\bfY,\underline{\widetilde\bfY},\underline\bfZ,\underline{\widetilde{\bfZ}}}{\text{maximize}}\quad&g_s(\underline\bfW,\underline\bGa,\underline\bfY,\underline{\widetilde\bfY},\underline\bfZ,\underline{\widetilde{\bfZ}})\\
\text{subject to}\quad &\eqref{cons:W}-\eqref{cons:Z}\\
&\widetilde{\bfZ}_{\ell k}\in\mathbb {C}^{N_{\ell}^r\times d_{\ell k}},
\end{align}
\end{subequations}
where the new objective function is shown in \eqref{obj:ISAC:GQT} as displayed at the top of the page with $\tilde\lambda_{\ell} =\lambda_{\text{max}}(\widetilde\bfL_{\ell})$. The above is the ultimately reformulated problem.

\begin{algorithm}[t]
  \caption{Nonhomogeneous FP for ISAC Beamforming}
  \label{algorithm:Nonhomogeneous FP}
  \begin{algorithmic}[1]
      \STATE Initialize $\underline{\bfW}$ to a feasible value, $\underline{\widetilde\bfY}$ according to \eqref{eq:update_tilde_Y}, and $\underline{\widetilde\bfZ}$ according to \eqref{eq:update tilde Z}.
      \REPEAT 
      \STATE Update each $\bfZ_{\ell k}$ by \eqref{eq:update Z}.
      \STATE Update each $\bm{\Gamma}_{\ell k}$ by \eqref{eq:update_Ga}.
      \STATE Update each $\bfY_{\ell k}$ and $\widetilde{\bfY}_{\ell k}$ by \eqref{eq:update_Y} and \eqref{eq:update tilde Y New}, respectively.
      \STATE Update each $\widetilde\bfZ_{\ell k}$ by \eqref{eq:update tilde Z}.
      \STATE Update each $\bfW_{\ell k}$ by \eqref{eq:projection_W}.
      \UNTIL{the objective value converges} 
  \end{algorithmic}
\end{algorithm}

Now, following the steps in Example \ref{example:1}, we consider optimizing the variables in \eqref{prob:ISAC gs} iteratively as
\begin{align*}
 \cdots\rightarrow\underline\bfW \rightarrow \underline\bfZ\rightarrow\underline\bGa\rightarrow\underline\bfY\rightarrow\underline{\widetilde\bfY} \rightarrow\underline{\widetilde\bfZ}\rightarrow \underline\bfW \rightarrow \cdots.   
\end{align*}
According to Lemma \ref{lemma:Nonhomo}, $\underline\bfZ$ and $\underline{\widetilde{\bfZ}}$ are optimally updated  as
\setcounter{equation}{57}
\begin{align}
    \bfZ_{\ell k}^\star &= \bfW_{\ell k},\label{eq:update tilde Z}\\
    \widetilde\bfZ_{\ell k}^\star &= \widetilde\bfY_{\ell k}. 
    \label{eq:update  Z}
\end{align}
When other variables are held fixed, $\underline\bGa$ is still optimally determined as in \eqref{eq:update_Ga} by solving $\partial g_s/\underline\bGa_{\ell}=\bm0$. When other variables are held fixed, $\underline{\bfY}$ is still optimally determined as in \eqref{eq:update_Y}; notice that updating $\bfY_{\ell k}$ requires computing the matrix inverse $\bfU_{\ell k}\in\mathbb C^{M_{\ell k}\times M_{\ell k}}$, but this is tolerable since the number of receiver antennas $M_{\ell k}$ is typically a small integer\footnote{If the number of receive antennas $M_{\ell k}$ is also a large number, then we need to apply the nonhomogeneous bound in Lemma \ref{lemma:Nonhomo} one more time so as to eliminate matrix inverse from the iterative update of $\bfY_{\ell k}$. We omit the detail of this extension since its idea is straightforward.}. With  the optimal $\underline{\widetilde\bfZ}=\underline{\widetilde\bfY}$ plugged in $g_s(\underline{\bfW},\underline\bGa,\underline\bfY,\underline{\widetilde{\bfY}},\underline{\bfZ},\underline{\widetilde\bfZ})$, we can find the optimal update of ${\bfW}_{\ell k}$ as
\begin{align}
    \bfW_{\ell k}^\star&=\arg\underset{\underline\bfW\in\mathcal{W}}{\min}\|\lambda_\ell\bfW_{\ell k}-\bm{\Lambda}_{\ell k}-(\lambda_{\ell}\bI-\bfL_{\ell})\bfZ_{\ell k})\|_F^2\notag\\
    &=\mathcal{P}_{\mathcal{W}}\left(\bfZ_{\ell k}+\frac{1}{\lambda_{\ell}}(\bm{\Lambda}_{\ell k}-\bfL_{\ell}\bfZ_{\ell k})\right).
    \label{opt W}
\end{align}
Equivalently, the optimal $\underline\bfW$ can be obtained as
\begin{equation}
\label{eq:projection_W}
\\\bfW^\star_{\ell k} =
\left\{ 
\begin{array}{ll}
    \!\!\widehat{\bfW}_{\ell k}  &\text{if}\;\sum^K_{j=1}\|\widehat\bfW_{\ell j}\|^2_F\le P_{\ell}
    \vspace{0.5em}\\
\!\!\sqrt{\frac{{P_{\ell}}}{\sum^K_{j=1}\|\widehat\bfW_{\ell j}\|^2_F}} \widehat\bfW_{\ell k}&\text{otherwise},
\end{array}
\right.
\end{equation}
where
\begin{align}
\label{eq:update_hat_W}
    \widehat{\bfW}_{\ell k} = \bfZ_{\ell k}+\frac{1}{\lambda_{\ell}}(\bm{\Lambda}_{\ell k}-\bfL_{\ell}\bfZ_{\ell k}).
\end{align}
After $\underline\bfZ$ has been optimally updated to $\underline{\bfW}$, we obtain the optimal update of $\underline{\widetilde\bfY}$ as
\begin{align}
     \widetilde\bfY_{\ell k}^\star&=\arg{\min}\|\tilde\lambda_\ell\widetilde\bfY_{\ell k}-\dot{\bfG}_{\ell\ell}\bfW_{\ell k}-(\tilde\lambda_{\ell}\bI-\widehat\bfQ_{\ell})\widetilde\bfZ_{\ell k})\|_F^2\notag\\
    &=\widetilde\bfZ_{\ell k}+\frac{1}{\tilde\lambda_{\ell}}(\dot{\bfG}_{\ell\ell}\bfW_{\ell k}-\widehat\bfQ_{\ell}\widetilde\bfZ_{\ell k}).
    \label{eq:update tilde Y New}
\end{align}
Algorithm \ref{algorithm:Nonhomogeneous FP} summarizes the above steps and is referred to as the nonhomogeneous FP method for the ISAC beamforming. Differing from the conventional FP algorithm in Algorithm \ref{algorithm:WMMSE}, the above new algorithm does not require computing any large matrix inverse, thanks to the nonhomogeneous bound in Lemma \ref{lemma:Nonhomo}.

Notice that Algorithm \ref{algorithm:Nonhomogeneous FP} is basically a composite application of the Lagrangian dual transform in Proposition \ref{prop:LDT}, the quadratic transform in Proposition \ref{prop:QT}, and the nonhomogeneous bound in Lemma \ref{lemma:Nonhomo}. In particular, it has been shown in \cite{shen2023convergence} that the above transformation techniques can all be interpreted as the MM procedure, so their combination, Algorithm \ref{algorithm:Nonhomogeneous FP}, is an MM method as well. As such, the convergence of Algorithm \ref{algorithm:Nonhomogeneous FP} can be immediately verified by the MM theory.
 
\begin{proposition}[Convergence Analysis]
\label{prop:convergence}
    Algorithm \ref{algorithm:Nonhomogeneous FP} can be interpreted as an MM method, i.e., $g_s(\underline{\bfW},\underline\bGa,\underline\bfY,\underline{\widetilde{\bfY}},\underline{\bfZ},\underline{\widetilde\bfZ})$ in    \eqref{obj:ISAC:GQT} amounts to a surrogate function of the original objective function in \eqref{obj:ISAC}. Thus, according to \cite{razaviyayn2013unified,sun2016majorization}, Algorithm \ref{algorithm:Nonhomogeneous FP} is guaranteed to converge to a stationary point of problem \eqref{prob:ISAC}, with the original objective value monotonically increasing after every iteration.
\end{proposition}
\begin{IEEEproof}
We prove convergence by using the MM theory---which is briefly reviewed in the following. For the original (nonconvex) problem 
\begin{align*}
    \underset{x\in\mathcal{X}}{\text{maximize}}\quad f(x),
\end{align*}
$g(x|\hat x)$ is said to be a surrogate function conditioned on $\hat x\in\mathcal{X}$ if the following two conditions are satisfied:
\begin{align}
\label{eq:inequ}
    g(x|\hat x)&\leq f(x) \quad\text{and}\quad
    g(\hat x|\hat x)= f(\hat x).
\end{align}
Then, instead of optimizing $x$ directly in the original problem, the MM method solves a sequence of new problems of the surrogate function:
\begin{align*}
    \underset{x\in\mathcal{X}}{\text{maximize}}\quad g(x|\hat x).
\end{align*}
As a main result from the MM theory, solving the above new problem iteratively guarantees convergence to a stationary point of the original problem so long as $f(x)$ and $g(x|\hat x)$ are differentiable.\\

In the ISAC problem case, the original objective function is defined in \eqref{prob:ISAC}, and the new objective function is defined in \eqref{obj:ISAC:GQT}. Now, 
treating $x = \underline{\bW}$ and $\hat{x}  = \{\underline\bGa,\underline\bfY,\underline{\widetilde{\bfY}},\underline{\bfZ},\underline{\widetilde\bfZ}\}$, we construct a surrogate function as
\begin{align*}
    g_n(\underline\bW|\underline{\widehat\bW}) =  g_s(\underline\bW,\underline{\bGa}(\underline{\widehat\bW}),\underline{\bY}(\underline{\widehat\bW}),\underline{\widetilde\bY}(\underline{\widehat\bW}),\underline{\bZ}(\underline{\widehat\bW}),\underline{\widetilde\bZ}(\underline{\widehat\bW})).
\end{align*}
It can be shown that
\begin{align*}
    g_n(\underline\bW|\underline{\widehat \bW})&\leq f_o(\underline\bW) \quad\text{and}\quad
    g_n(\underline{\widehat \bW}|\underline{\widehat \bW})= f_o(\widehat \bW).
\end{align*}
Since the original ISAC objective function and the reformulated objective function are differentiable, the convergence to a stationary point of the original problem can be guaranteed according to the MM theory.
\end{IEEEproof}

We further compare the per-iteration complexities of Algorithm \ref{algorithm:WMMSE} and Algorithm \ref{algorithm:Nonhomogeneous FP} in what follows. To render the complexity analysis tractable, we assume that every $N^t_{\ell}=N_t$, every $N^r_{\ell}=N_r$, and every $d_{\ell k}=d$. We first focus on the update of each ${\bfW}_{\ell k}$. For Algorithm \ref{algorithm:WMMSE}, the main complexity comes from \eqref{eq:update_W} and \eqref{eq:update_eta}, which is $\mathcal{O}(tN_t^3+td^2N_t)=\mathcal{O}(N_t^3)$, where $t$ is the bisection iteration that is needed to find the optimal $\eta_{\ell}$. The main complexity of the proposed method comes from {the computation of $\lambda_{\ell}$},  \eqref{eq:projection_W} and \eqref{eq:update_hat_W}. {If $\lambda_{\ell} = \lambda_{\max}(\bL_{\ell})$}, we employ the power method \cite{datta2004numerical} for eigenvalue computation. Its complexity is $\mathcal{O}(cN^2_t)=\mathcal{O}(N_t^2)$, where $c$ is the number of iterations required. The overall complexity of the two updates \eqref{eq:projection_W} and \eqref{eq:update_hat_W} equals $\mathcal{O}(dN_t^2+d^2N_t)$. Therefore, the complexity of our proposed method is $\mathcal{O}(N_t^2)$.  Recall that there are two other possible choices for $\lambda_{\ell}$ as stated in Remark \ref{remark:lambda}: $\lambda_{\ell} = \tr(\bL_{\ell})$ requires a complexity of $\mathcal{O}(N_t)$, and $\lambda_{\ell} = \|\bL_{\ell}\|_F$ requires a complexity of $\mathcal{O}(N^2_t)$. Nevertheless, the above two substitutes for $\lambda_{\ell} = \lambda_{\max}(\bL_{\ell})$ do not impact the overall complexity of the algorithm, since the main complexity comes from the computations in \eqref{eq:projection_W} and  \eqref{eq:update_hat_W}. The complexity of the proposed method is much lower than the conventional method since in practice $d$ is much smaller than $N_t$. Likewise, when updating $\underline{\widetilde{\bfY}}$, the complexity of the conventional method is $\mathcal{O}(N_r^3)$, while the complexity of the proposed method is only  $\mathcal{O}(N_r^2).$

\subsection{Proposed Fast FP for ISAC Beamforming}
\label{subsec:Fast FP}

In this work, we not only aim to eliminate large matrix inverse to reduce the per-iteration complexity (as already done in Section \ref{nonhomogeneous ISAC}), but also seek to accelerate the convergence in iterations. Our approach is based upon the following crucial observation: Algorithm \ref{algorithm:Nonhomogeneous FP} has a deep connection with gradient projection, so Nesterov's extrapolation strategy comes into play.

\begin{figure}[t]
\centering
\includegraphics[width=1.0\linewidth]{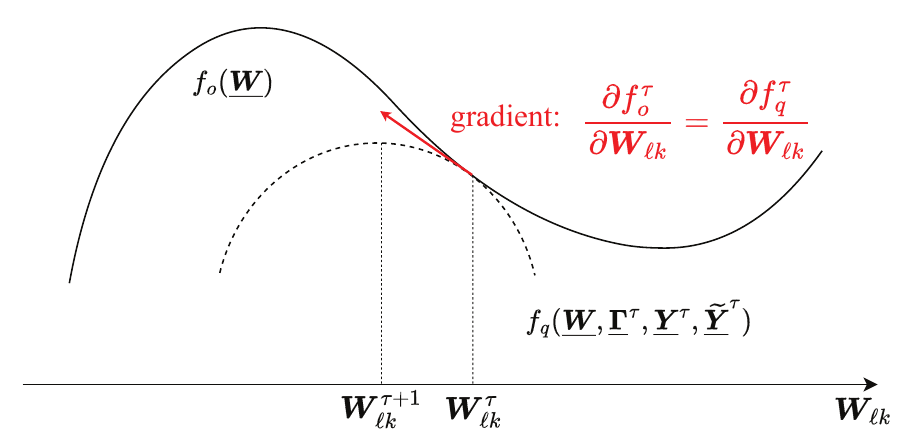}
\caption{In the $\tau$th iteration, $f_q$ and $f_o$ have the same gradient with respect to each $\bfW_{\ell k}$ after the updates of the auxiliary variables $(\underline\bGa,\underline\bfY,\underline{\widetilde\bfY})$.}
\label{fig:MM}
\end{figure}

First and foremost, we describe the connection between Algorithm \ref{algorithm:Nonhomogeneous FP} and gradient projection in the following proposition.
\begin{proposition}
\label{prop:proj}
Algorithm \ref{algorithm:Nonhomogeneous FP} is equivalent to a gradient projection, i.e., updating the primal variable $\bfW_{\ell k}$ iteratively as in \eqref{eq:projection_W} amounts to
\begin{equation}
  \bfW^{\tau+1}_{\ell k}  = \mathcal{P}_{\mathcal W}\bigg(\bfW^{\tau}_{\ell k} + \zeta^\tau\cdot\frac{\partial f_o(\underline\bfW^{\tau})}{\partial \bfW_{\ell k}}\bigg),
\end{equation}
where $\tau$ is the iteration index, $\zeta^\tau>0$ is the gradient step size in the $\tau$th iteration, and $f_o(\underline\bfW^{\tau})$ is the primal objective function \eqref{obj:ISAC}.
\end{proposition}
\begin{IEEEproof}
It is difficult to compute the partial derivative $\partial f_o(\underline\bfW^{\tau})/\partial \bfW_{\ell k}$ directly. We suggest taking advantage of a property of the surrogate function. Recall that $f_q$ in \eqref{obj:ISAC:QT} is a surrogate function of the original objective function $f_o$. According to the MM theory, $f_o\ge f_q$ where the equality holds only after one round of the auxiliary variable updates has been finished. In other words, the gap between $f_o$ and $f_q$ is minimized to be zero at this point, so we have
\begin{equation}
    \frac{\partial (f_o^\tau-f_q^\tau)}{\partial \bfW_{\ell k}}=\bm 0,\;\forall (\ell,k) 
\end{equation}
in the $\tau$th iteration, where the superscript $\tau$ is the iteration index. Thus, as illustrated in Fig.~\ref{fig:MM}, we further have
\begin{align}
   \label{f_o:gradient}
    \frac{\partial f_o^{\tau}}{\partial \bfW_{\ell k}}&=\frac{\partial f_q^{\tau}}{\partial \bfW_{\ell k}}.
\end{align}
Importantly, it turns out that the partial derivative of $f_q$ can be easily obtained as
\begin{align}
    \frac{\partial f_o^{\tau}}{\partial \bfW_{\ell k}}&=\frac{\partial f_q(\underline{\bfW}^{\tau},\underline\bGa^{\tau},\underline\bfY^{\tau},\underline{\widetilde{\bfY}}^{\tau})}{\partial \bfW_{\ell k}}\notag\\
    &=\bm{\Lambda}^{\tau}_{\ell k}-\bfL^{\tau}_{\ell}\bfW^{\tau}_{\ell k},
    \label{projection proof}
\end{align}
where $\bm{\Lambda}_{\ell k}$ and $\bfL_\ell$ are defined as in \eqref{matrix:Lambda} and \eqref{matrix L} with $(\underline{\bGa}^\tau,\underline\bfY^{\tau},\underline{\widetilde{\bfY}}^\tau)$.

We then shift our attention to the iterative update of $\bfW_{\ell k}$ in \eqref{opt W} by Algorithm \ref{algorithm:Nonhomogeneous FP}:
\begin{align}
    \bfW^{\tau+1}_{\ell k} &= \mathcal{P}_{\mathcal W}\bigg(\bfZ^\tau_{\ell k}+\frac{1}{\lambda^\tau_{\ell}}(\bm{\Lambda}^\tau_{\ell k}-\bfL^\tau_{\ell}\bfZ^\tau_{\ell k})\bigg)\notag\\
    &\overset{(a)}{=} \mathcal{P}_{\mathcal W}\bigg(\bfW^{\tau}_{\ell k} + \frac{1}{\lambda^{\tau}_\ell}\Big(\bm{\Lambda}^\tau_{\ell k}-\bfL^\tau_{\ell}\bfW^{\tau}_{\ell k}\Big)\bigg)\notag\\
    &\overset{(b)}{=} \mathcal{P}_{\mathcal W}\bigg(\bfW^{\tau}_{\ell k} + \frac{1}{\lambda^{\tau}_\ell}\cdot\frac{\partial f_o(\underline\bfW^{\tau})}{\partial \bfW_{\ell k}}\bigg),\notag
\end{align}
where step $(a)$ follows by the optimal update of $\bfZ_{\ell k}$ in \eqref{eq:update tilde Z}, and step $(b)$ follows by the combination of \eqref{f_o:gradient} and \eqref{projection proof}. The proof is then completed.     
\end{IEEEproof}
\begin{remark}
Although the previous works \cite{zhang2023enhancing,shen2023convergence} already explored the connection between FP and gradient projection, they only consider eliminating the matrix inverse for the iterative update of $\bfW_{\ell k}$, with the nonhomogenous bound applied only once. In other words, if we stop at $g_o(\underline\bfW,\underline\bGa,\underline\bfY,\underline{\widetilde\bfY},{{\bfZ}})$ in \eqref{prob:ISAC go} and consider optimizing these variables iteratively, then we would recover the results in \cite{shen2023convergence}. In contrast, Algorithm \ref{algorithm:Nonhomogeneous FP} applies the nonhomogeneous bound twice in order to eliminate the matrix inverse for both $\bfW_{\ell k}$ and $\widetilde\bfY_{\ell k}$. Surprisingly, we have shown that the connection with gradient projection continues to hold after two uses of the nonhomogeneous bound. Moreover, we remark that it is difficult to carry over the relevant proofs in \cite{shen2023convergence} and \cite{zhang2023enhancing} to our problem case directly. The proof of Proposition \ref{prop:proj} is based on a new idea.
\end{remark}

\begin{figure}[t]
\centering
\includegraphics[width=0.9\linewidth]{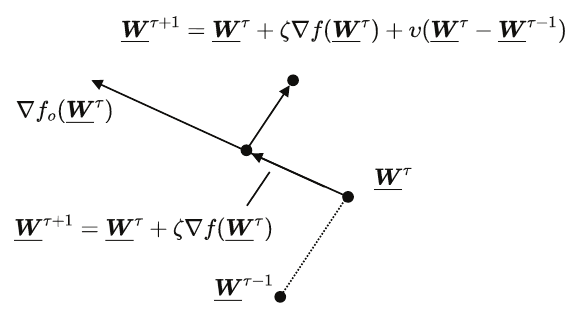}
\caption{Nesterov's gradient ascent with extrapolation \cite{Nesterov_book}.}
\label{fig:extrapolation}
\end{figure}

\begin{algorithm}[t]
  \caption{Fast FP for ISAC Beamforming}
  \label{algorithm:Fast FP}
  \begin{algorithmic}[1]
      \STATE Initialize $\underline{\bfW}$ to a feasible value, $\underline{\widetilde\bfY}$ according to \eqref{eq:update_tilde_Y}, and $\underline{\widetilde\bfZ}$ according to \eqref{eq:update tilde Z}.
      \REPEAT 
      \STATE Update each ${\bfV}_{\ell k}$ according to \eqref{extrapolate} and set ${\bfW}_{\ell k}={\bfV}_{\ell k}$.
      \STATE Update each $\bfZ_{\ell k}$ by \eqref{eq:update Z}.
      \STATE Update each $\bm{\Gamma}_{\ell k}$ by \eqref{eq:update_Ga}.
      \STATE Update each $\bfY_{\ell k}$ and $\widetilde{\bfY}_{\ell k}$ by \eqref{eq:update_Y} and \eqref{eq:update tilde Y New}, respectively.
      \STATE Update each $\widetilde\bfZ_{\ell k}$ by \eqref{eq:update tilde Z}.
      \STATE Update each $\bfW_{\ell k}$ by \eqref{eq:projection_W}.
      \UNTIL{the objective value converges} 
  \end{algorithmic}
\end{algorithm}

In light of the connection between Algorithm \ref{algorithm:Nonhomogeneous FP} and gradient projection, we can readily accelerate Algorithm \ref{algorithm:Nonhomogeneous FP} by using Nesterov's extrapolation strategy \cite{Nesterov_book}---which was initially designed for the gradient method. As illustrated in Fig.~\ref{fig:extrapolation}, we now extrapolate each $\bfW_{\ell k}$ along the direction of the difference between the preceding two iterates before the gradient projection, i.e.,
\begin{align}
    {\bfV}^{\tau-1}_{\ell k} &= \bfW_{\ell k}^{\tau-1}+\upsilon^{\tau-1}(\bfW_{\ell k}^{\tau-1}-\bfW_{\ell k}^{\tau-2}),\label{extrapolate}\\
    \bfW_{\ell k}^{\tau} &=  \mathcal{P}_{\mathcal{W}}\left({\bfV}^{\tau-1}_{\ell k}+\frac{1}{\lambda^\tau_{\ell}}(\bm{\Lambda}_{\ell k}-\bfL_{\ell}{\bfV}^{\tau-1}_{\ell k})\right),
\end{align}
where $\tau$ is the iteration index,  the extrapolation step $\upsilon_{\tau}$ is chosen as
\begin{equation}
    \upsilon^\tau = \max\bigg\{\frac{\tau-2}{\tau+1},0\bigg\},\;\;\text{for}\;\tau=1,2,\ldots,
\end{equation}
the starting point is $\underline{\bfW}^{-1}=\underline{\bfW}^0$. 
Algorithm \ref{algorithm:Fast FP} summarizes the above steps and is referred to as the fast FP algorithm.
\begin{remark}
    The fast FP method is not limited to the ISAC problem considered in this paper. It can be applied to the general weighted-sum-of-traces-of-matrix-ratio problem:
\begin{align}
    \underset{\underline\bX}{\text{maximize}}\quad \sum_{i=1}^n \omega_i\tr(\bC_i\bM_i(\underline\bX)),
\end{align}
where $\bC_i\in\mathbb{H}_{++}^{d\times d}$ is a positive definite coefficient matrix. The problem with the $\log\big|\bm I+\bC_i\bM_i(\underline\bX)\big|$ term is also doable since we can convert that term to $\tr(\bC_i\bM_i(\underline\bX))$ by the Lagrangian dual transform as stated in Proposition \ref{prop:LDT}.
\end{remark}

\begin{figure*}[t]
\centering
\subfigure[The ISAC objective value in \eqref{prob:ISAC} v.s. iteration number ]
{
\includegraphics[width=0.48\linewidth]{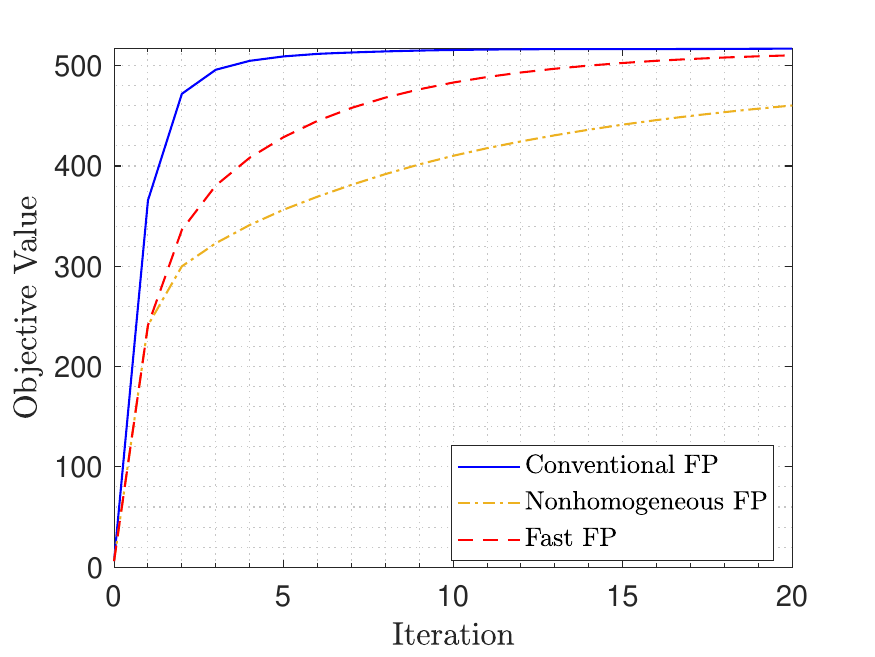}
}
\subfigure[The ISAC objective value v.s. elapsed time]
{
\includegraphics[width=0.48\linewidth]{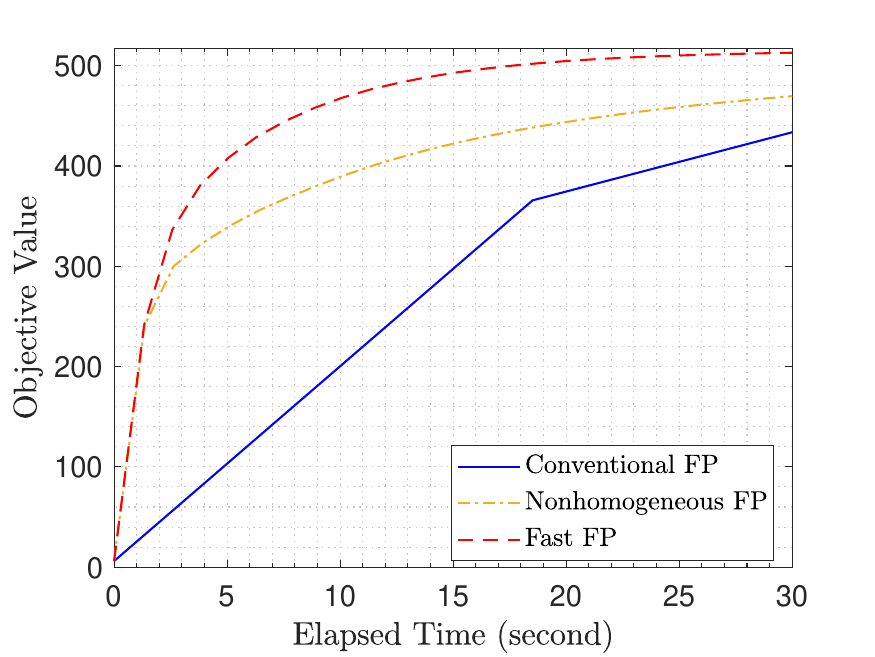}
}
\caption{{The convergence behaviors of the different ISAC beamforming algorithms when $N_r=N_t=128$.} }
\label{fig:convergence:128}
\end{figure*}
\begin{figure*}[ht]
\centering
\subfigure[The ISAC objective value v.s. iteration number]
{
\includegraphics[width=0.48\linewidth]{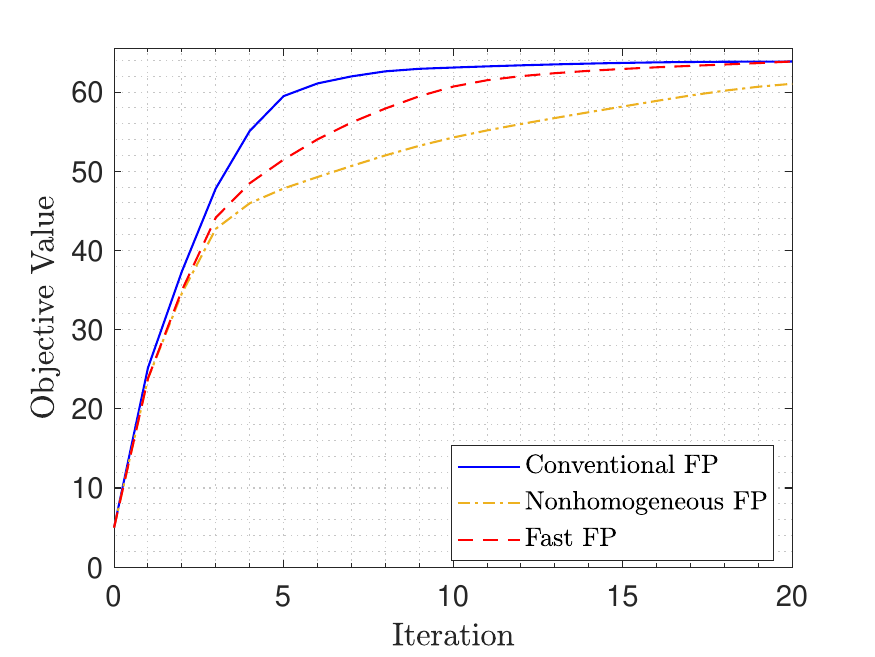}
}
\subfigure[The ISAC objective value v.s. elapsed time]
{
\includegraphics[width=0.48\linewidth]{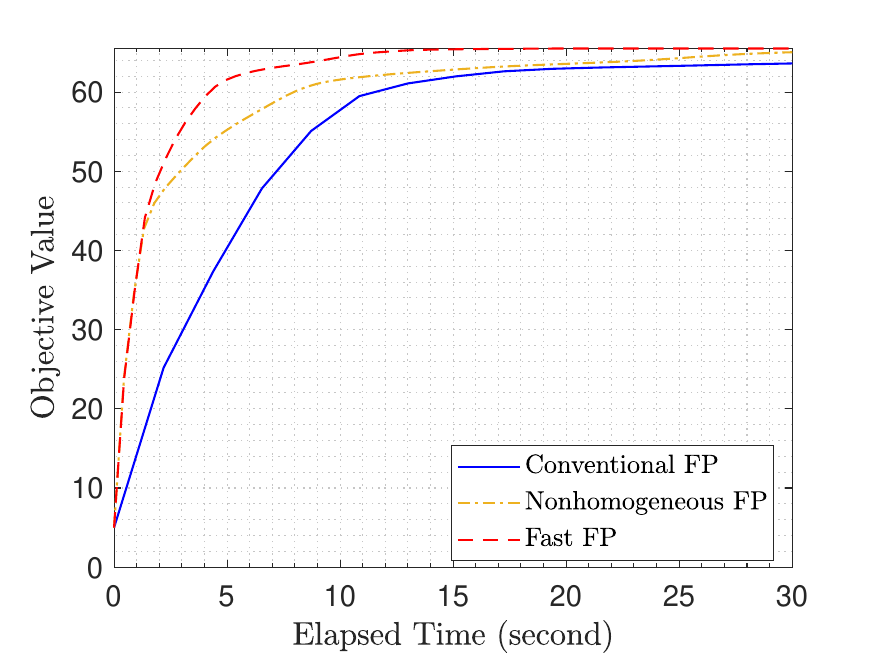}
}
\caption{{The convergence behaviors of the different ISAC beamforming algorithms when $N_r=N_t=4$.}}
\label{fig:convergence:4}
\end{figure*}

In light of the MM theory \cite{shen2023convergence}, we can characterize the convergence rates of the different algorithms.
Let $f_o(\cdot)$ be the optimization objective value, let $\underline{\bW}^\star$ be a stationary point that the algorithm converges to, let $k$ be the iterate index, let $R$ be the Euclidean distance from the starting point to $\underline{\bW}^\star$, let $L$ be the Lipschitz constant of $\nabla^2 f_o(\underline{\bW})$, let $\Lambda$ be the maximum eigenvalue of $\nabla^2 (f_o(\underline{\bW})-g_s(\underline{\bW},\underline{\bGa},\underline{\bY},\underline{\widetilde\bY},\underline{\bZ},\underline{\widetilde\bZ}))$, and let $\underline{\bW}^{k}$ be the solution after $k$ iterates. The convergence rate of Algorithms \ref{algorithm:Nonhomogeneous FP} and Algorithm \ref{algorithm:Fast FP} are given in the following proposition.
\begin{proposition}[Convergence Rate]
    The local convergence rate of Algorithm \ref{algorithm:Nonhomogeneous FP} is 
    \begin{align}
        f_o(\underline{\bW}^\star)-f_o(\underline{\bW}^1)&\leq \frac{\Lambda R^2}{2}+\frac{LR^3}{6},\\
        f_o(\underline{\bW}^\star)-f_o(\underline{\bW}^k)&\leq \frac{2\Lambda R^2+2LR^3/3}{k+3},\; \text{for }k\ge2.
    \end{align}
    The local convergence rate of Algorithm \ref{algorithm:Fast FP} is
        \begin{align}
        f_o(\underline{\bW}^\star)-f_o(\underline{\bW}^k)\leq \frac{2LR_0}{(k+1)^2},\; \text{for }k\ge 1,
    \end{align}
    where $R_0>0$ is a constant depending on the starting point.
\end{proposition}

The proof of the above proposition follows the convergence analysis in \cite{shen2023convergence}, both based on the MM theory.

\section{Numerical Results}
\label{sec:simulation}

\begin{figure*}[ht]
\centering
\subfigure[The ISAC objective value v.s. iteration number]
{
\includegraphics[width=0.48\linewidth]{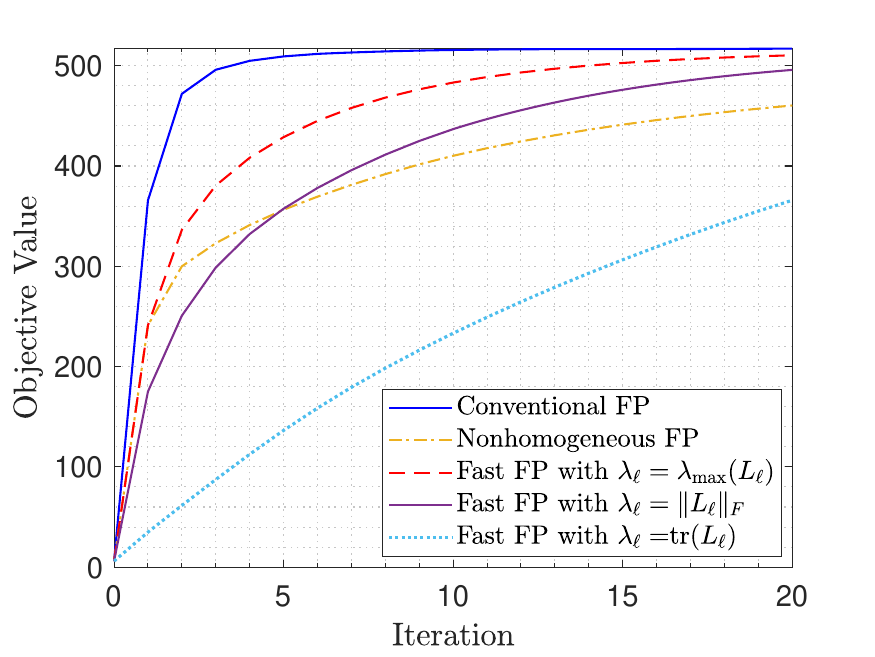}
}
\subfigure[The ISAC objective value v.s. elapsed time]
{
\includegraphics[width=0.48\linewidth]{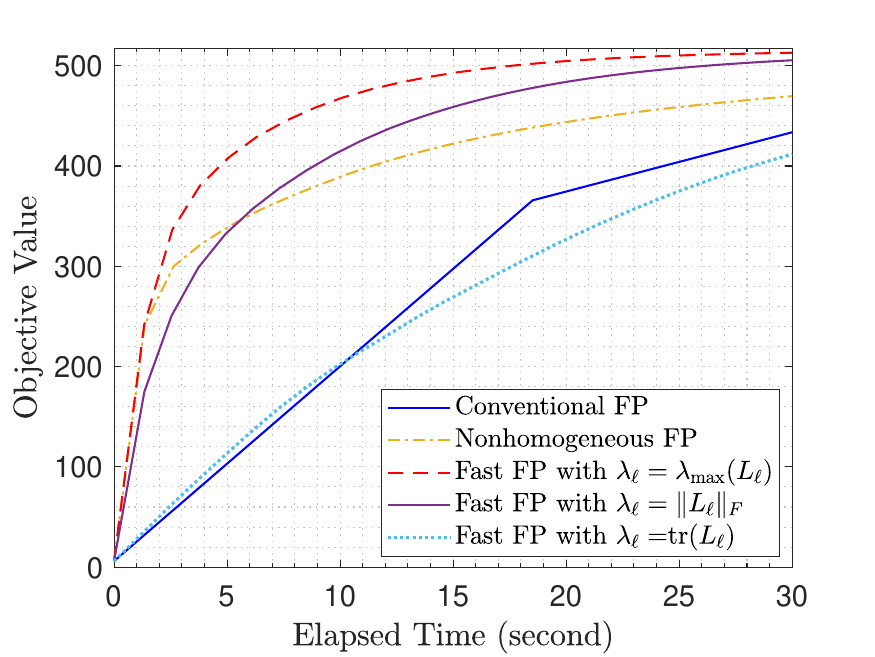}
}
\caption{The convergence of the ISAC objective value with the different values of $\lambda_{\ell}$ for the fast FP method in Algorithm \ref{algorithm:Fast FP}.}
\label{fig:convergence:different lambda}
\end{figure*}

We validate the performance of the proposed algorithms numerically in a 7-cell wrapped-hexagonal-around network. Within each cell, the BS is located at the center, with the BS-to-BS distance being $800$ meters,  while its associated downlink users are randomly distributed; we make the downlink users be close to the cell edge on purpose in order to stress the cross-cell interference effect. Let every BS have the same number of transmit (resp. receive) antennas denoted by $N_t$ (resp. $N_r$). Let every downlink user have $4$ antennas;  $4$ data streams are intended for each of them. The maximum transmit power $P_\ell$ at each BS is set to $20$ dBm, the background noise power level at the downlink user side $\sigma^2$ is $-80$ dBm, and the background noise power level at the BS side $\tilde\sigma^2$ is set to {$-70$} dBm. The distance-dependent path-loss is given by $15.3 + 37.6 \log_{10}(d) + \kappa $ (in dB), where $d$ represents the BS-to-user distance in meters, and $\kappa$ is a zero-mean Gaussian random variable with a standard variance of $8$ dB---which models the shadowing effect. For the sensing, every reflection coefficient $\xi_{\ell}$ is set to {$10^{-3}$} as in {\cite{hua2023mimo}}. The block length $T$ equals $30$. We use the same starting point for all the competitor algorithms for comparison fairness.

\subsection{Rate of Convergence}
\label{simulation: Convergence Comparison}
We begin with the convergence speed performance of the various methods in solving problem \eqref{prob:ISAC}. Assume that there are $45$ downlink users in each cell. The priority weights in problem \eqref{prob:ISAC} are set as $\beta_{\ell} = {10^{-14}}$  and $\omega_{\ell k} = 1$. (Notice that $\beta_{\ell}$ is considerably smaller because the Fisher information value is usually much greater than the data rate value.) { The parameter $\lambda_{\ell}$ is set to $\lambda_{\ell} = \lambda_{\max}(\bL_{\ell})$ for nonhomogeneous FP and fast FP by default.}

Fig.~\ref{fig:convergence:128} shows the convergence behaviors of the different algorithms when $N_t=N_r=128$, i.e., when both transmit antennas and radar antennas are massively deployed. Observe from Fig.~\ref{fig:convergence:128}(a) that the conventional FP converges faster than the nonhomogeneous FP and the fast FP in iterations. From an MM perspective, this is because the conventional FP gives a tighter approximation of the original objective function $f_o(\underline{\bfW})$. However, this does not imply that the absolute running time of the conventional FP is the shortest, because in the meanwhile the conventional FP requires heavier computations per iteration due to the large matrix inversion. Actually, it can be seen from Fig.~\ref{fig:convergence:128}(b) that the nonhomogeneous FP converges faster than the conventional FP in time. For example, the conventional FP requires about $19$ seconds to reach the objective value of $366$, while the nonhomogeneous FP merely requires $8$ seconds. Thus, in practice, the nonhomogeneous FP still runs much faster than the conventional FP. Further, observe also that the fast FP outperforms the nonhomogeneous FP not only in terms of the iteration efficiency but also in terms of the time efficiency. For example, to reach the objective value of 380, the nonhomogeneous FP requires $7$ iterations and $9$ seconds, while the fast FP just requires $3$ iterations and $3.8$ seconds. Thus, the fast FP can further improve upon the nonhomogeneous FP significantly, thanks to Nesterov's acceleration strategy. Moreover, we consider in Fig.~\ref{fig:convergence:4} the small-antenna-array case of $N_t=N_r=4$. According to the figure, although the advantage of the fast FP over the nonhomogeneous FP now becomes smaller, it still outperforms the conventional FP significantly in terms of running time.

We now compare the convergence rate of the fast FP when its key parameter $\lambda_{\ell}$ is set to different values. We first consider the convergence rate in terms of iteration as shown in Fig.~\ref{fig:convergence:different lambda}(a). Observe that $\lambda_\ell=\lambda_{\max}(\bm L_\ell)$ has the fastest convergence because it gives the tightest surrogate function for the original objective. In contrast, $\lambda_\ell = \mathrm{tr}(\bm L_\ell)$ leads to the loosest surrogate function and thus has the slowest convergence. There is a similar observation in Fig.~\ref{fig:convergence:different lambda}(b) when the convergence is considered in terms of elapsed time.

\begin{table*}[t]
\footnotesize
\renewcommand{\arraystretch}{1.4}
  \centering
  \caption{{Estimation of $\theta_\ell$ by the different algorithms under the running time limit of 5 seconds.}}
    \begin{tabular}{l||rrrrr|rrrrr}
     \hline
        & \multicolumn{5}{c|}{Mean Squared Error} & \multicolumn{5}{c}{Max Squared Error } \\
    \hline
    Target Position & \multicolumn{1}{c}{a} & \multicolumn{1}{c}{b} & \multicolumn{1}{c}{c} & \multicolumn{1}{c}{d} & \multicolumn{1}{c|}{e} & \multicolumn{1}{c}{a} & \multicolumn{1}{c}{b} & \multicolumn{1}{c}{c} & \multicolumn{1}{c}{d} & \multicolumn{1}{c}{e} \\  
    \hline
    \hline
    Rough DoA & 2.33E$-3$ & 2.34E$-3$ & 2.89E$-3$ & 3.80E$-3$ & 1.96E$-3$ & 2.33E$-3$ & 2.34E$-3$ & 2.89E$-3$ & 3.80E$-3$ & 1.96E$-3$ \\
    Conventional FP &3.73E$-5$ & 5.13E$-5$ & 1.22E$-4$ & 3.73E$-5$ & 1.66E$-4$ & 3.89E$-5$ & 8.10E$-5$ & 3.73E$-4$ & 3.89E$-5$ & 4.07E$-4$ \\
    Nonhomogeneous FP & 3.73E$-5$ & 4.93E$-5$ & 1.07E$-4$ & 3.65E$-5$ & 1.53E$-4$ & 3.89E$-5$ & 7.06E$-5$ & 1.53E$-4$ & 3.89E$-5$ & 3.73E$-4$ \\
    Proposed Fast FP   & 3.73E$-5$ & 4.93E$-5$ & 9.88E$-5$ & 3.50E$-5$ & 9.93E$-5$ & 3.89E$-5$ & 7.06E$-5$ & 1.53E$-4$ & 3.50E$-5$ & 1.42E$-4$ \\
    \hline
    \end{tabular}%
  \label{tab:est}%
\end{table*}%

\subsection{Angle Estimation Performance}
\begin{figure}[t]
    \centering
    \includegraphics[width=1.0\linewidth]{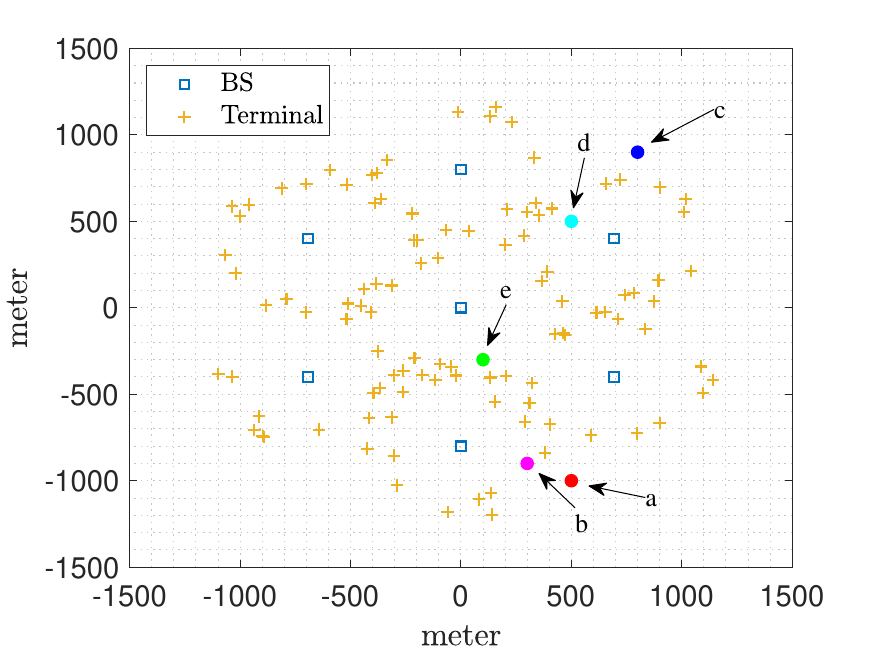}
    \caption{Positions of BSs, downlink users, and the sensing targets in a 7-cell wrapped-around ISAC network.}
    \label{fig:Location}
\end{figure}

Next, we focus on the sensing performance by considering the estimation of $\theta_\ell$ under the different ISAC beamforming algorithms. The transmit power of each BS is fixed at $10$ dBm. Let $N_t=N_r=128$; let every reflection coefficient $\xi_{\ell}=10^{-4}$; let each $\beta_\ell=10^{-8}$ and let each $\omega_{\ell k}=1$; we now adopt a smaller reflection coefficient $\xi_{\ell}$ than in Section \ref{simulation: Convergence Comparison} to test the sensing performance of different algorithms under poor conditions and a larger weight $\beta_\ell$ for sensing to highlight the sensing performance. Now assume that each cell contains a total of $15$ downlink users. Consider the following 5 possible target positions as depicted in Fig.~\ref{fig:Location}:
\begin{itemize}
    \item Position a: $(500\text{m},-1000\text{m})$
    \item Position b: $(300\text{m},-900\text{m})$
    \item Position c: $(800\text{m},900\text{m})$
    \item Position d: $(500\text{m},500\text{m})$
    \item Position e: $(100\text{m},-300\text{m})$
\end{itemize}
We shall sequentially place the target at the above positions.

The angle estimation method comes from \cite{song2023intelligent}, as briefly described in the following for completeness. Let $\bfX_{\ell}$ be the transmit signal of BS $\ell$, i.e., $\bfX_{\ell}=\sum_{k}\bfW_{\ell k}\bfS_{\ell k}$. The $\theta_{\ell}$ estimate is given by
\begin{align}
\label{eq:est_theta}
    \hat{\theta}_\ell=\arg \underset{\theta_\ell}{\max}\; \frac{|\text{tr}(\bfG_{\ell,\ell}(\theta_\ell)^\text{H} \widetilde{\bm{\Psi}}_{\ell} \bfX_{\ell}^\text{H})|^2}{\ttr((\bfG_{\ell \ell}(\theta_\ell)\bfX_{\ell})^\bhh(\bfG_{\ell \ell}(\theta_\ell)\bfX_{\ell}))}.
\end{align}
In particular, to account for time efficiency, we carry out each algorithm under a time-limited setting, i.e., the running time is limited to 5 seconds. Since the BSs do not know the exact location of the target, we design the beamformer based on their rough knowledge of DoA; the resulting rough knowledge error is shown in Table~\ref{tab:est}. Regarding the estimation error metric, we consider two choices: (i) the mean squared error $\frac{1}{L}\sum_{\ell}|\hat\theta_{\ell}-\theta_{\ell}|^2$ and (ii) the max squared error $\max_{\ell} |\hat\theta_{\ell}-\theta_{\ell}|^2$. The performances of the different algorithms are summarized in Table~\ref{tab:est}. Observe from Table~\ref{tab:est} that the conventional FP
has the worst performance. For example, at position c, the nonhomogeneous FP reduces the mean squared error by about 7.83\% as compared to the conventional FP, while the fast FP reduces the mean squared error by about 40.18\%.
Thus, although all the algorithms can guarantee convergence to the stationary point, the conventional FP is much more time-consuming and thus is inferior to the other two methods when the running time is limited. It is worth noticing that the nonhomogeneous FP and the fast FP end up with the same estimation error for some target positions, but the latter achieves a higher objective value than the former. The reason is that the mean squared error and the max squared error are sensitive to the specific estimation algorithm; clearly, the estimation algorithm \cite{song2023intelligent} we adopt does not differ much between the nonhomogeneous FP and the fast FP. In contrast, according to the general metric of the Fisher information that is independent of the specific estimation algorithm, the fast FP still outperforms the nonhomogeneous FP, as shown in the previous simulation results. 
\subsection{Communications v.s. Sensing}

\begin{figure}[t]
    \centering
    \includegraphics[width=1.0\linewidth]{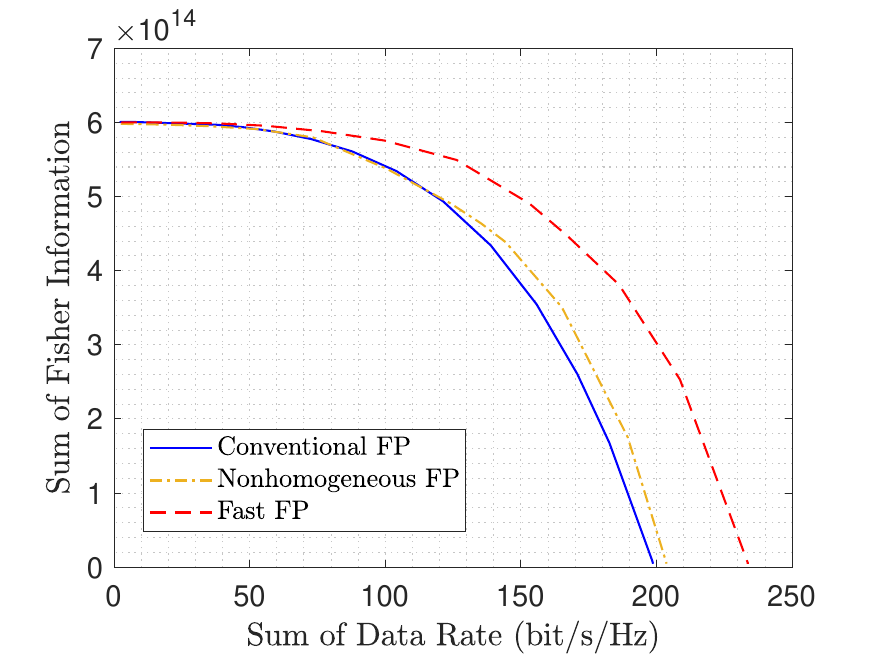}
    \caption{{Tradeoff between the sum of rates and the sum of Fisher information by varying the value of $\omega_{\ell k}$ with each $\beta_{\ell}$ fixed at $10^{-14}$.}}
    \label{fig:tradeoff_timelimited}
\end{figure}

\begin{figure}[t]
    \centering
    \includegraphics[width=1.0\linewidth]{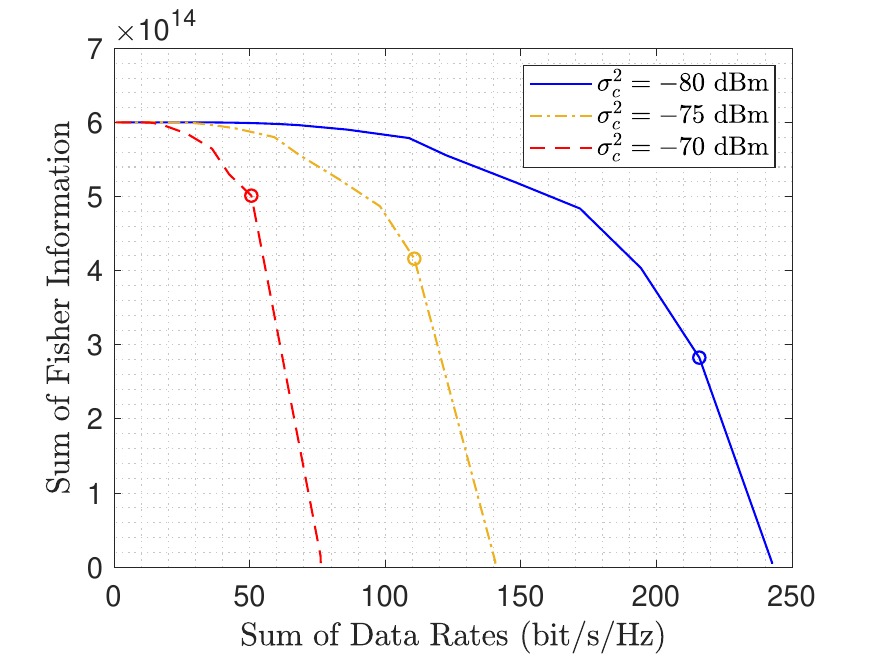}
    \caption{Tradeoff between the sum of rates and the sum of Fisher information by varying the value of $\omega_{\ell k}$ while fixing $\beta_{\ell}$ at $10^{-14}$. The circled points stand for the case with each $\omega_{
    \ell k}=0.1$.}
    \label{fig:tradeoff2}
\end{figure}
\begin{figure*}[ht]
\centering
\subfigure[Sum of data rates v.s. $N_t$]
{
\includegraphics[width=0.48\linewidth]{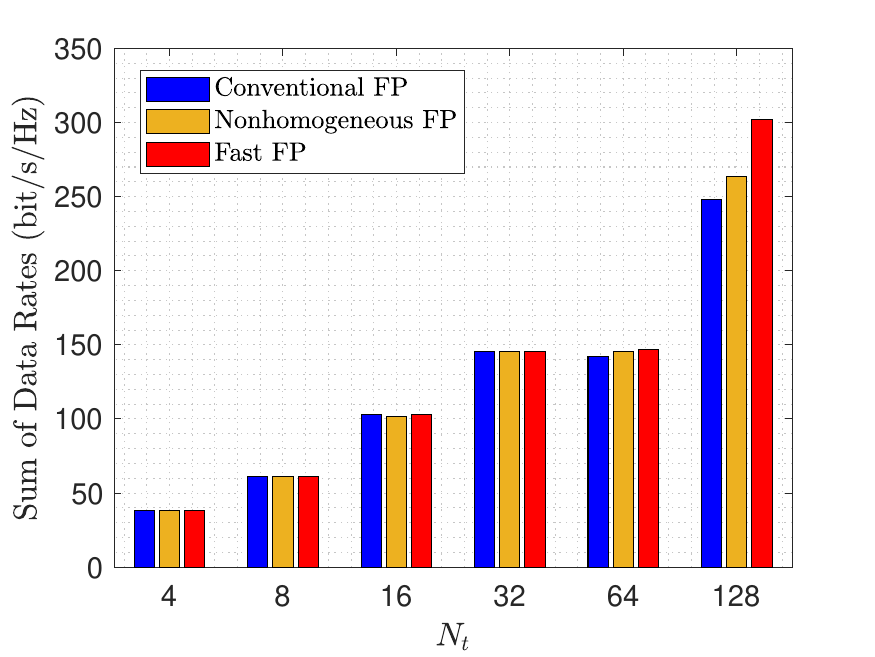}
}
\subfigure[Sum of Fisher information v.s. $N_t$]
{
\includegraphics[width=0.48\linewidth]{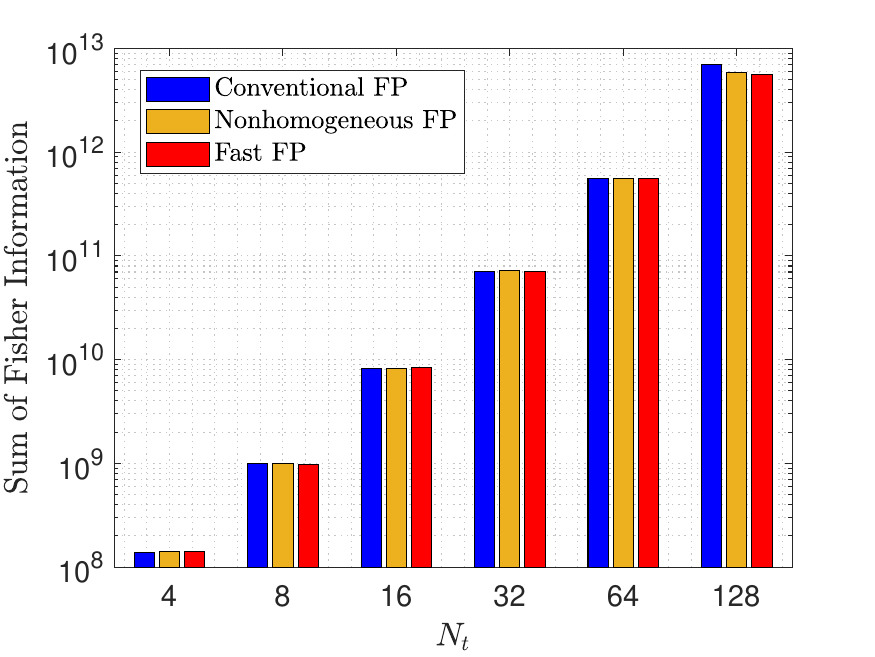}
}
\caption{{ The sum of data rates v.s. the sum of Fisher information v.s. $N_t$ under the running time limit of 5 seconds.}}
\label{fig:Nt:limited time}
\end{figure*}

\begin{figure}[t]
\centering
\includegraphics[width=1.0\linewidth]{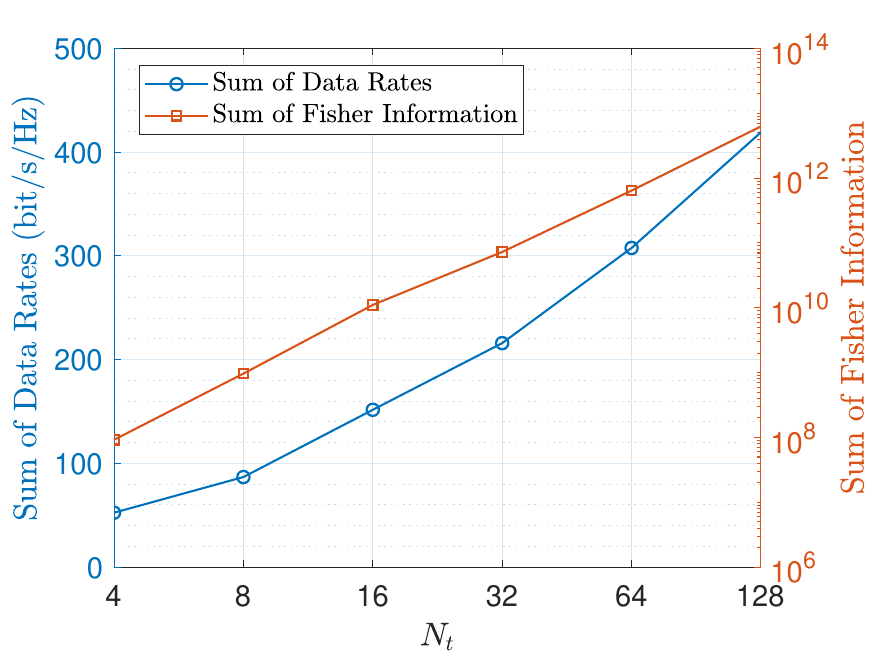}
    \caption{{The sum of data rates vs. the sum of Fisher information vs. $N_t$.}}
    \label{fig:different Nt}
\end{figure}

\begin{figure}[t]
    \centering
    \includegraphics[width=1.0\linewidth]{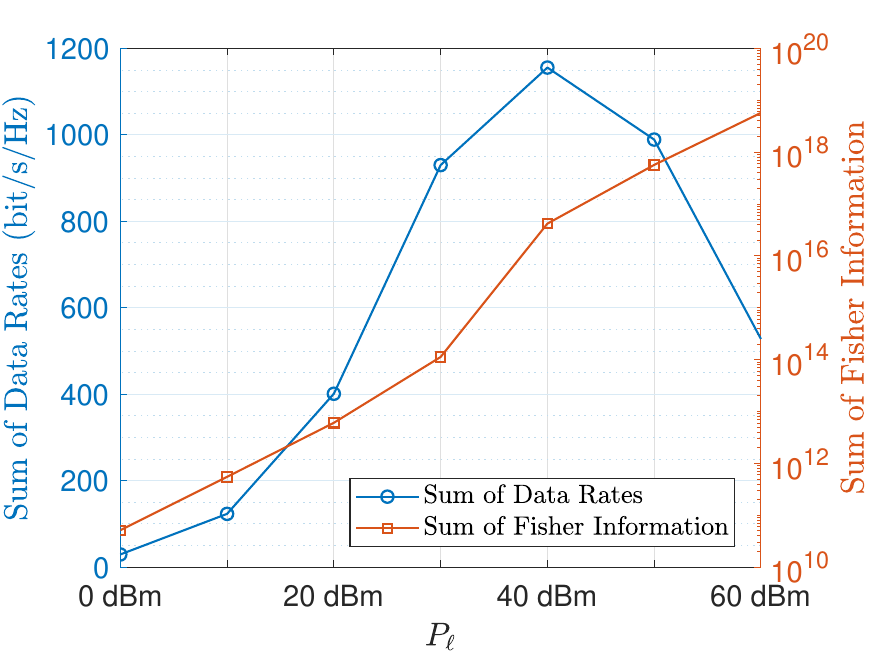}
    \caption{{The sum of data rates vs. the sum of Fisher information vs. $P_{\ell}$.}}
    \label{fig:different P}
\end{figure}

Lastly, we examine the tradeoff between the communications performance and the sensing performance. We now assume that there are 15 downlink users in each cell. To enable tradeoff, we fix the sensing weights $\beta_\ell$ at $10^{-14}$ while varying the communications weights $\omega_{\ell k}$ ranging from $10^{-10}$ to $10^{10}$. First, we consider a time-limited setting, i.e., the running time is limited to $6$ seconds. Observe from Fig.~\ref{fig:tradeoff_timelimited} that the tradeoff curve of the fast FP lies outside the tradeoff curves of the conventional FP and the nonhomogeneous FP, so its performance is superior. For example, if the sum of Fisher information is fixed at $3\times10^{14}$, the fast FP can enhance the sum of data rates by $17.65$\% as compared to the nonhomogeneous FP, and by $21.21$\% as compared to the conventional FP.

Now we drop the running time limit and allow each algorithm to run till convergence. Fig.~\ref{fig:tradeoff2} shows the tradeoff curve between the sum of data rates and the sum of Fisher information under the different noisy environments. Notice that the tradeoff curves meet at the same point when $\omega_{\ell k}$ tends to $10^{-10}$ so that the beamforming is considered for sensing alone. When we give more weight to communications, the three curves would diverge. The figure shows that the sum of data rates increases most rapidly with $\omega_{\ell k}$ raised to 0.1 when $\sigma^2=-80$ dBm, i.e., in the high-SNR regime. Moreover, regardless of the noise power level $\sigma^2$, as the communications weights get higher, we always observe that the growth of data rates is fast at the beginning and then slows down.

Furthermore, we look into the ISAC objective more closely by comparing the sum of data rates and the sum of Fisher information separately for the different choices of the number of transmit antennas $N_t$. When the running time limit of 5 seconds is imposed on each algorithm, observe from Fig.~\ref{fig:Nt:limited time}(a) that the three methods yield similar performance when $N_t$ is small; this makes sense because all the algorithms can fully converge to the stationary points when the per-iteration complexity is low. However, when $N_t$ increases, the number of iteration decreases because of the increasing per-iteration complexity, so that the algorithms can no longer attain convergence. Clearly, the conventional FP is most sensitive to the increased complexity. As shown in Fig.~\ref{fig:Nt:limited time}(a), the conventional FP is inferior to the other methods when $N_t$ becomes large. In the meanwhile, the sensing performance metric is less impacted by the increased $N_t$, as shown in Fig.~\ref{fig:Nt:limited time}(b). The conventional FP even attains a higher sum of Fisher information when $N_t=128$.

Fig.~\ref{fig:different Nt} plots the sum of data rates versus the sum of Fisher information without the running time limit for the algorithms.  Notice that the sum of data rates enhances with $N_t$ linearly at most of the time. For example, when $N_t$ increases from 8 to 16, the sum of data rates increases by around $50$\%. In comparison, the improvement of the Fisher information is much sharper. For example, when $N_t$ increases from 8 to 16, the sum of the Fisher information is enhanced by almost an order of magnitude. Thus, our method is capable of reaping the DoF gain when the large antenna arrays are deployed.

Furthermore, we plot the sum of data rates and the sum of Fisher information separately for the different choices of the power budget $P_\ell$ in Fig.~\ref{fig:different P}. The communication performance and the sensing performance are both increasing with $P_\ell$ when $P_\ell < 40 $ dBm; however, when $P_\ell > 40 $ dBm, the communication performance starts to decrease while the sensing performance continues to increase. The reason is that the payoff of the Fisher information is much higher than that of the data rate when power is sufficiently high; this makes sense since the data rate is a concave function of the power. To avoid overemphasizing the sensing performance when $P_\ell$ becomes higher, we need to reduce the weight $\beta_\ell$ accordingly.       
\section{Conclusion}
\label{sec:conclusion}

This paper considers the ISAC beamforming optimization for a multi-cell massive MIMO network. Under the FP framework, we first show that the classic WMMSE algorithm works for the ISAC case, but at a high computation cost because it requires inverting large matrices. We then develop a nonhomogeneous approximation to eliminate the large matrix inversion from the iterative algorithm. Furthermore, we connect the above algorithm to gradient projection, and thereby employ Nesterov's extrapolation scheme to accelerate the convergence.

\label{sec:refs}
\bibliographystyle{IEEEtran}
\bibliography{IEEEabrv,refs}

\vskip 0pt plus -1fil

\begin{IEEEbiographynophoto}
{Yannan Chen} received the B.E. degree in Automation Engineering and the M.E. degree in Pattern Recognition and Intelligent Systems from Xiamen University, Xiamen, China, in 2018 and 2021, respectively. He is currently pursuing his Ph.D. degree with the School of Science and Engineering at The Chinese University of Hong Kong (Shenzhen). His research interests include optimization, wireless communications, and machine learning.
\end{IEEEbiographynophoto}

\vskip 0pt plus -1fil

\begin{IEEEbiographynophoto}
{Yi Feng} received the B.E. degree in Mathematics and Applied Mathematics from Shanghai Jiao Tong University, Shanghai, China, in 2023. He is currently pursuing his Ph.D. degree with the School of Science and Engineering at The Chinese University of Hong Kong (Shenzhen). His research interests include optimization, wireless communications, and information theory.
\end{IEEEbiographynophoto}

\vskip 0pt plus -1fil

\begin{IEEEbiographynophoto}
{Xiaoyang Li} is currently a Research Associate with the Shenzhen Research Institute of Big Data, and an adjunct Assistant Professor with the Chinese University of Hong Kong, Shenzhen. He received the Ph.D. degree from The University of Hong Kong. His research interests include integrated sensing-communication-computation and network optimization. He is a recipient of the Young Elite of China Association for Science and Technology, the “Forbes China 30 under 30”, the Young Elite of G20 Entrepreneurs’ Alliance, the Best Paper Award from IEEE JC\&S, and the Exemplary Reviewers of IEEE WCL and JII. He served as the editor of JII and workshop chairs of IEEE ICASSP/WCNC/PIMRC/MIIS.
\end{IEEEbiographynophoto}

\vskip 0pt plus -1fil

\begin{IEEEbiographynophoto}
{Licheng Zhao} received the B.S. degree in Information Engineering from Southeast University (SEU), Nanjing, China, in 2014, and the Ph.D. degree with the Department of Electronic and Computer Engineering at the Hong Kong University of Science and Technology (HKUST), in 2018. Since June 2018, he has been an algorithm engineer in recommendation system with JD.COM, China. Since Dec. 2021, he has served as a research scientist in Shenzhen Research Institute of Big Data (SRIBD).  His research interests are in optimization theory and efficient algorithms, with applications in signal processing, machine learning, and deep learning in recommendation system.
\end{IEEEbiographynophoto}

\vskip 0pt plus -1fil

\begin{IEEEbiographynophoto}
{Kaiming Shen} received the B.Eng. degree in information security and the B.Sc. degree in mathematics from Shanghai Jiao Tong University, China in 2011, and then the M.A.Sc. degree in electrical and computer engineering from the University of Toronto, Canada in 2013. After working at a tech startup in Ottawa for one year, he returned to the University of Toronto and received the Ph.D. degree in electrical and computer engineering in early 2020. Dr. Shen has been with the School of Science and Engineering at The Chinese University of Hong Kong (CUHK), Shenzhen, China as a tenure-track assistant professor since 2020. His research interests include optimization, wireless communications, information theory, and machine learning.
Dr. Shen received the IEEE Signal Processing Society Young Author Best Paper Award in 2021, the CUHK Teaching Achievement Award in 2023, and the Frontiers of Science Award at the International Congress of Basic Science in 2024. Dr. Shen currently serves as an Editor for IEEE Transactions on Wireless Communications.
\end{IEEEbiographynophoto}

\end{document}